\documentclass[a4paper,11pt]{article}
\pdfoutput=1
\usepackage{jcappub} % for details on the use of the package, please see the JINST-author-manual
\usepackage{lineno}
\usepackage[T1]{fontenc} % if needed
\usepackage{subcaption}
\usepackage{array}

\usepackage{graphicx}% Include figure files
\usepackage{dcolumn}% Align table columns on the decimal point
\usepackage{bm}% bold math
\usepackage{xspace}
\usepackage{xcolor}
\usepackage{hyperref}% add hypertext capabilities
\hypersetup{colorlinks=true,citecolor=blue,filecolor=blue,urlcolor=blue,}
\usepackage[noabbrev]{cleveref}
\usepackage{orcidlink}

\defcitealias{DESI2024.IV.KP6}{\texttt{DESI2024-\lya}}
\newcommand{\kplya}{\citetalias{DESI2024.IV.KP6}\xspace}

\newcommand{\lya}{Ly$\alpha$}
\newcommand{\lyaf}{Ly$\alpha$ forest}
\newcommand{\lyb}{Ly$\beta$}

\newcommand{\lyacolore}{\texttt{Ly$\alpha$CoLoRe}}
\newcommand{\saclay}{\texttt{Saclay}}
\newcommand{\iminuit}{\texttt{iminuit}}

\newcommand{\apar}{$\alpha_{||}$}
\newcommand{\atrans}{$\alpha_\bot$}

\newcommand{\lyalyalyalya}{Ly$\alpha$(A)$\times$Ly$\alpha$(A)}
\newcommand{\lyalyalyalyb}{Ly$\alpha$(A)$\times$Ly$\alpha$(B)}
\newcommand{\lyalyaqso}{Ly$\alpha$(A)$\times$QSO}
\newcommand{\lyalybqso}{Ly$\alpha$(B)$\times$QSO}

\newcommand{\hMpc}{\ $h^{-1}\text{Mpc}$}

\newcommand{\picca}{\texttt{picca}}
\newcommand{\vega}{\texttt{Vega}}

\renewcommand{\thefootnote}{\fnsymbol{footnote}}

\title{Validation of the DESI 2024 \lyaf\ BAO analysis using synthetic datasets}
\author[1]{Andrei Cuceu\orcidlink{0000-0002-2169-0595},\footnote[2]{NASA Einstein Fellow}}
\author[2]{{Hiram K.~Herrera-Alcantar}\orcidlink{0000-0002-9136-9609},}
\author[3]{{Calum Gordon},}
\author[1,4]{{Paul Martini}\orcidlink{0000-0002-4279-4182},}
\author[5]{{Julien Guy},}
\author[3]{{Andreu Font-Ribera}\orcidlink{0000-0002-3033-7312},}
\author[6,2]{{Alma X.~Gonzalez-Morales}\orcidlink{0000-0003-4089-6924},}
\author[7]{{M.~Abdul Karim}\orcidlink{0009-0000-7133-142X},}
\author[5]{{J.~Aguilar},}
\author[8]{{S.~Ahlen}\orcidlink{0000-0001-6098-7247},}
\author[7]{{E.~Armengaud}\orcidlink{0000-0001-7600-5148},}
\author[9]{{A.~Bault}\orcidlink{0000-0002-9964-1005},}
\author[10]{{D.~Brooks},}
\author[5]{{T.~Claybaugh},}
\author[11]{{A.~de la Macorra}\orcidlink{0000-0002-1769-1640},}
\author[10]{{P.~Doel},}
\author[12,13]{{K.~Fanning}\orcidlink{0000-0003-2371-3356},}
\author[5,14]{{S.~Ferraro}\orcidlink{0000-0003-4992-7854},}
\author[15,16]{{J.~E.~Forero-Romero}\orcidlink{0000-0002-2890-3725},}
\author[17,18,19]{{E.~Gaztañaga},}
\author[5]{{S.~Gontcho A Gontcho}\orcidlink{0000-0003-3142-233X},}
\author[20]{{G.~Gutierrez},}
\author[1,21]{{K.~Honscheid},}
\author[22]{{C.~Howlett}\orcidlink{0000-0002-1081-9410},}
\author[1,4,21]{{N.~G.~Kara{\c c}ayl{\i}}\orcidlink{0000-0001-7336-8912},}
\author[9]{{D.~Kirkby}\orcidlink{0000-0002-8828-5463},}
\author[5]{{A.~Kremin}\orcidlink{0000-0001-6356-7424},}
\author[5]{{M.~Landriau}\orcidlink{0000-0003-1838-8528},}
\author[7]{{J.M.~Le~Goff},}
\author[23]{{L.~Le~Guillou}\orcidlink{0000-0001-7178-8868},}
\author[5]{{M.~E.~Levi}\orcidlink{0000-0003-1887-1018},}
\author[24,3]{{M.~Manera}\orcidlink{0000-0003-4962-8934},}
\author[25]{{A.~Meisner}\orcidlink{0000-0002-1125-7384},}
\author[26,3]{{R.~Miquel},}
\author[27]{{J.~Moustakas}\orcidlink{0000-0002-2733-4559},}
\author[11]{{A.~Muñoz-Gutiérrez},}
\author[28]{{A.~D.~Myers},}
\author[2,29]{{G.~Niz}\orcidlink{0000-0002-1544-8946},}
\author[7,5]{{N.~Palanque-Delabrouille}\orcidlink{0000-0003-3188-784X},}
\author[30,31,32]{{W.~J.~Percival}\orcidlink{0000-0002-0644-5727},}
\author[5,33,14]{{C.~Poppett},}
\author[34]{{F.~Prada}\orcidlink{0000-0001-7145-8674},}
\author[35]{{I.~P\'erez-R\`afols}\orcidlink{0000-0001-6979-0125},}
\author[3]{{C.~Ram\'irez-P\'erez},}
\author[36,7,37]{{C.~Ravoux}\orcidlink{0000-0002-3500-6635},}
\author[38]{{M.~Rezaie}\orcidlink{0000-0001-5589-7116},}
\author[39]{{G.~Rossi},}
\author[40]{{E.~Sanchez}\orcidlink{0000-0002-9646-8198},}
\author[5]{{D.~Schlegel},}
\author[41,42]{{M.~Schubnell},}
\author[43]{{H.~Seo}\orcidlink{0000-0002-6588-3508},}
\author[25]{{D.~Sprayberry},}
\author[7]{{T.~Tan}\orcidlink{0000-0001-8289-1481},}
\author[42]{{G.~Tarl\'{e}}\orcidlink{0000-0003-1704-0781},}
\author[11]{{M.~Vargas-Maga\~na}\orcidlink{0000-0003-3841-1836},}
\author[44,45]{{M.~Walther}\orcidlink{0000-0002-1748-3745},}
\author[25]{{B.~A.~Weaver},}
\author[5]{{R.~Zhou}\orcidlink{0000-0001-5381-4372},}
\author[46]{{H.~Zou}\orcidlink{0000-0002-6684-3997}}

\affiliation{Affiliations are in Appendix \ref{sec:affiliations}}

\emailAdd{cuceu.1@osu.edu}

\abstract{
The first year of data from the Dark Energy Spectroscopic Instrument (DESI) contains the largest set of Lyman-$\alpha$ (\lya) forest spectra ever observed. This data, collected in the DESI Data Release 1 (DR1) sample, has been used to measure the Baryon Acoustic Oscillation (BAO) feature at redshift $z=2.33$. In this work, we use a set of 150 synthetic realizations of DESI DR1 to validate the DESI 2024 \lyaf\ BAO measurement presented in \cite{DESI2024.IV.KP6}. The synthetic data sets are based on Gaussian random fields using the log-normal approximation. We produce realistic synthetic DESI spectra that include all major contaminants affecting the \lyaf. The synthetic data sets span a redshift range $1.8<z<3.8$, and are analysed using the same framework and pipeline used for the DESI 2024 \lyaf\ BAO measurement. To measure BAO, we use both the \lya\ auto-correlation and its cross-correlation with quasar positions. We use the mean of correlation functions from the set of DESI DR1 realizations to show that our model is able to recover unbiased measurements of the BAO position. We also fit each mock individually and study the population of BAO fits in order to validate BAO uncertainties and test our method for estimating the covariance matrix of the \lyaf\ correlation functions. Finally, we discuss the implications of our results and identify the needs for the next generation of \lyaf\ synthetic data sets, with the top priority being to simulate the effect of BAO broadening due to non-linear evolution.
}

\begin{document}
\maketitle
\flushbottom
\renewcommand{\thefootnote}{\arabic{footnote}}

%%%%%%%%%%%%%%%%%%%%%%%%%%%%%%%%%%%%%%%%%%%%%%%%%%%%%%%%%%%%%%%%%%%%%%%%%%%%%%%%%%%%%%%%%%%%%%%%5

\section{Introduction}
\label{sec:intro}

Baryon Acoustic Oscillations (BAO) measured from large-scale structure surveys have been extensively used to map cosmic expansion across the history of the Universe \cite{Eisenstein:2005,Cole:2005,Beutler:2011,Ross:2015,Alam:2017,Abbott:2019,Alam:2021,Abbott:2022,DES:2024}, providing some of the tightest cosmological constraints to date \cite[e.g.,][]{Alam:2017,Alam:2021}. The ongoing Dark Energy Spectroscopic Instrument (DESI, \cite{DESI2016a.Science,DESI2016b.Instr,DESI2022.KP1.Instr,DESI2023a.KP1.SV,DESI2023b.KP1.EDR}) survey aims to map an order of magnitude more galaxies and quasars compared to previous spectroscopic surveys, in order to obtain the next generation of BAO constraints across a wide range of redshifts ($0<z<4$). DESI finished collecting the first year of data in June 2022, and this first year data assembly (hereafter DESI DR1) contains roughly 13 million galaxies and 1.5 million quasars over $9500$ square degrees. The DESI DR1 sample is presented in \cite{DESI2024.I.DR1}, and has been used to measure BAO from the distribution of galaxies at redshifts $z<2$, presented in \cite{DESI2024.II.KP3,DESI2024.III.KP4}, and using the Lyman-$\alpha$ (\lya) forest at redshifts $z>2$, presented in \cite{DESI2024.IV.KP6} (hereafter \kplya). The cosmological constraints from all DESI DR1 BAO measurements are presented in \cite{DESI2024.VI.KP7A}.

The BAO feature has been measured using the \lyaf\ for more than a decade now. The first measurements used the auto-correlation of \lya\ flux overdensities \cite{Busca:2013,Slosar:2013,Kirkby:2013} from the Baryon Oscillation Spectroscopic Survey (BOSS, \cite{Dawson2013}). Soon after, the cross-correlation between the \lyaf\ and quasars was also used to measure the BAO feature \cite{FontRibera2014}. Subsequent BOSS and extended BOSS (eBOSS, \cite{Dawson2016}) analyses improved on these measurements with larger datasets and better analysis and modelling tools \cite{Delubac:2015,Bautista:2017,duMasdesBourboux:2017,deSainteAgathe:2019,Blomqvist:2019,duMasdesBourboux:2020}. The final eBOSS \lyaf\ BAO analysis was presented in \cite{duMasdesBourboux:2020}, and constituted the state-of-the-art \lya\ BAO measurement until the first DESI measurement (\kplya).

This publication presents the validation of the DESI DR1 \lyaf\ BAO measurement from \kplya using synthetic data sets (mocks). We aim to use simulated DESI DR1 \lyaf\ data sets containing all major contaminants to test for potential systematic errors that could affect the measurement. We will also use a large set of mocks to stress test the analysis pipeline, study estimates of the covariance matrix, and understand the population of potential \lya\ BAO constraints from DESI DR1. The work here was performed in parallel with the measurement in \kplya, and our results were used to inform decisions for the analysis of DESI data.

To validate the DESI DR1 \lya\ BAO measurement, we generate synthetic realizations of the high-redshift part of DESI DR1 ($z>1.8$). These synthetic data sets are based on a Gaussian random field, with quasar positions drawn from its log-normal transformation. The \lya\ transmitted flux is computed from the Gaussian field along skewers to each quasar using the fluctuating Gunn-Peterson approximation \cite[FGPA;][]{Bi:1997,Croft:1998}. In this work, we use two different types of mocks based on this method \cite{Ramirez-Perez:2022,Farr:2020,Etourneau:2023}. The algorithms behind each set of mocks have been used before to generate mocks for the validation of the final \lyaf\ eBOSS analysis in \cite{duMasdesBourboux:2020}, and are described below in \Cref{sec:mocks}. We use these methods to generate 150 synthetic realizations of DESI DR1, each of them containing a simulated quasar catalogue, and also simulated spectra containing the \lyaf\ for each quasar. The process for generating the simulated spectra was introduced in \cite{Herrera-Alcantar:2024}, and is also described below in \Cref{sec:mocks}.

We analyze the set of 150 mocks with the same method and pipeline as was used for the real data in \kplya. The analysis and modelling process is described in \Cref{sec:analysis}. We present our results in \Cref{sec:results}, where we perform two types of analysis. For the first one, we combine the information from all mocks to obtain very high statistics correlation function measurements which we use to validate \lyaf\ BAO constraints with unprecedented precision. For the second, we perform the analysis individually for each mock and study the resulting population of BAO constraints. We discuss the implications of our work for \kplya and future DESI \lyaf\ analyses in \Cref{sec:discussion}, and summarize in \Cref{sec:conclusions}.

%%%%%%%%%%%%%%%%%%%%%%%%%%%%%%%%%%%%%%%%%%%%%%%%%%%%%%%%%%%%%%%%%%%%%%%%%%%%%%%%%%%%%%%%%%%%%%%%5

\section{DESI DR1 Ly$\alpha$ forest synthetic datasets}
\label{sec:mocks}

The process we use to make synthetic realizations of the DESI DR1 data set closely follows that used for DESI EDR mocks, which was presented in detail in \cite{Herrera-Alcantar:2024}. Therefore, we only give a summary of this process here, focusing on the differences with respect to the DESI EDR mocks. The mock creation process is broken into two steps. The first step involves drawing a Gaussian random field to simulate the matter density field, then using the log-normal transformation of this field to draw quasar positions, and finally simulating skewers of \lya\ transmitted flux along the line of sight to each quasar. We give a summary of this process in \Cref{subsec:mock_boxes}. The second step involves simulating realistic quasar populations that mimic the DESI DR1 survey properties (\Cref{subsec:mock_qso}) and turning skewers of \lya\ transmitted flux into realistic DESI spectra (\Cref{subsec:mock_spec}). One important change in our mocks compared to previous iterations involves re-tuning the absorption strength of metals present in the \lyaf\ to better match real data. We describe this in \Cref{subsec:metal_tune}.

\subsection{Transmitted flux boxes}
\label{subsec:mock_boxes}

The first step in our process of making synthetic realizations of the DESI DR1 data set involves using Gaussian random fields to simulate matter-density light cones. These matter density fields are then used to draw quasar positions using the log-normal approximation and to simulate \lya\ transmitted flux skewers to each quasar. We discuss our use of log-normal mocks along with their limitations in \Cref{sec:discussion}. Similarly to \cite{Herrera-Alcantar:2024}, we use two types of mocks produced by two different methods. We will refer to these as \lyacolore\ and \saclay\ mocks.

\lyacolore\ mocks were created through a two-step process (see \cite{Ramirez-Perez:2022} and \cite{Farr:2020} for detailed descriptions of the two steps). First, the \texttt{CoLoRe} package\footnote{\url{https://github.com/damonge/CoLoRe}} \cite{Ramirez-Perez:2022} was used to create low-resolution Gaussian random fields in $\sim(10 \;h^{-1}\text{Gpc})^3$ boxes, simulating a light cone to redshift $z=3.8$. The quasar positions are then drawn by Poisson sampling the log-normal transformation of the density field using an input number density and bias. The radial velocity field is computed from the gradient of the Newtonian potential and used together with the initial density field to simulate line-of-sight skewers from each quasar to the centre of the box \cite{Ramirez-Perez:2022}. The second step involves using the \lyacolore\ package\footnote{\url{https://github.com/igmhub/LyaCoLoRe}} \cite{Farr:2020} to post-process the skewers generated by \texttt{CoLoRe} into realistic \lya\ transmitted flux skewers. As the resolution used so far ($\sim2.4$ \hMpc) is too low for simulating small-scale \lyaf\ fluctuations, \lyacolore\ adds an extra one-dimensional Gaussian random field to each line-of-sight, which is based on measurements of the one-dimensional \lyaf\ power spectrum \cite{Farr:2020}.\footnote{This extra one-dimensional power does not affect the three-dimensional clustering because we do not use pixels in the same forest when computing correlation functions.} The log-normal transformation of the resulting field is used to compute the optical depth $\tau$ using the fluctuating Gunn-Peterson approximation \cite{Bi:1997,Croft:1998}, and redshift space distortions (RSD) are added using the radial velocity field from \texttt{CoLoRe}. Finally, the transmitted flux fraction is given by $F=e^{-\tau}$. \lyacolore\ also simulates a population of damped \lya\ absorbers (DLAs) with the same method used to draw quasars. These will be added as contaminants to our spectra later on (\Cref{subsec:mock_spec}).

\saclay\ mocks were created using the \texttt{SaclayMocks} package,\footnote{\url{https://github.com/igmhub/SaclayMocks}} which is described in \cite{Etourneau:2023}. This process is similar to the one above, but it makes use of multiple boxes to more accurately simulate the quasar distribution and the velocity field. For drawing quasar positions, the method employed by \texttt{CoLoRe} works well on large scales, but on small scales, it results in a quasar auto-correlation function that is significantly larger than the linear prediction or observations \cite{Laurent:2017}. To address this issue, instead of drawing quasars from the log-normal transformation of the matter density field, \cite{Etourneau:2023} uses a separate quasar-density box which is produced by modifying the original Gaussian random field in Fourier space to simulate a quasar-density field by accounting for the quasar bias. \saclay\ mocks also use dedicated boxes produced from the same underlying density field to simulate the radial peculiar velocity field and its gradient. A modified form of FGPA that accounts for the line-of-sight velocity gradient is then used to simulate the \lya\ transmitted flux fraction \cite{Etourneau:2023}. Finally, these mocks also simulate the distribution of DLAs.

For this work, we use a set of 100 \lyacolore\ boxes and 50 \saclay\ boxes. We are limited to these numbers mainly by computational and storage constraints. Note that we use the same set of \lyacolore\ boxes as \cite{duMasdesBourboux:2020,Cuceu:2023,Herrera-Alcantar:2024}. The boxes are produced with quasar number densities of $\sim120\;\text{deg}^{-2}$ for \lyacolore\ mocks, and $\sim100\;\text{deg}^{-2}$ for \saclay\ mocks, which are significantly larger than the DESI value of $\sim60\;\text{deg}^{-2}$. The resolution is also similar between the two, with $\sim2.4$\hMpc\ for \lyacolore\ mocks, and $\sim2.19$\hMpc\ for \saclay\ mocks.

The cosmologies used to produce the two sets of mocks are slightly different, as \lyacolore\ mocks are based on the Planck 2015 results \cite[Column 1 of Table 3 in][]{Planck:2016}, while \saclay\ mocks are based on the Planck 2018 results \cite[Column 5 of Table 1 in][]{Planck:2020}. However, the difference between these two is negligible for our purposes ($\sim0.02\%$ change in the BAO position), so we use the Planck 2015 cosmology to analyse all the mocks,\footnote{This choice is based on the fact that we have more \lyacolore\ boxes.} and we account for the small difference in the BAO results from \saclay\ mocks by moving them to the correct cosmology in post-processing (see \Cref{subsec:model}).

%%%%%%%%%%%%%%%%%%%%%%%%%%%%%%%%%%%%%%%%%%%%%%%%%%%%%%%%%%%%%%%%%%%%%%%%%%%%%%%%%%%%%%%%%%%%%%%%5

\subsection{Simulating DESI DR1 quasar populations}\label{subsec:mock_qso}

\begin{figure}
\centering
\begin{subfigure}{.66\textwidth}
    \centering
    \includegraphics[width=1.0\textwidth,keepaspectratio]{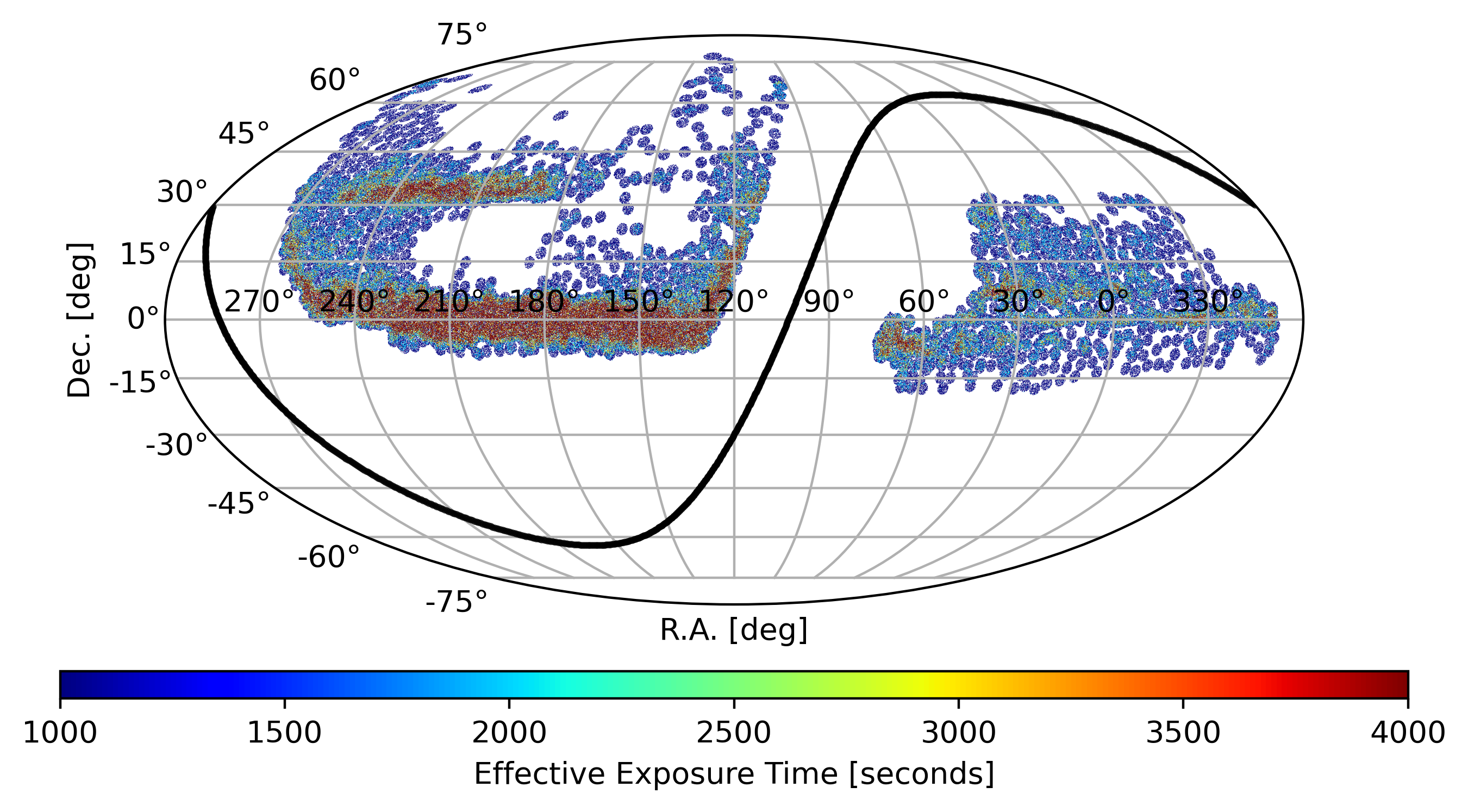}
\end{subfigure}
\begin{subfigure}{.33\textwidth}
    \centering
    \includegraphics[width=1.0\textwidth,keepaspectratio]{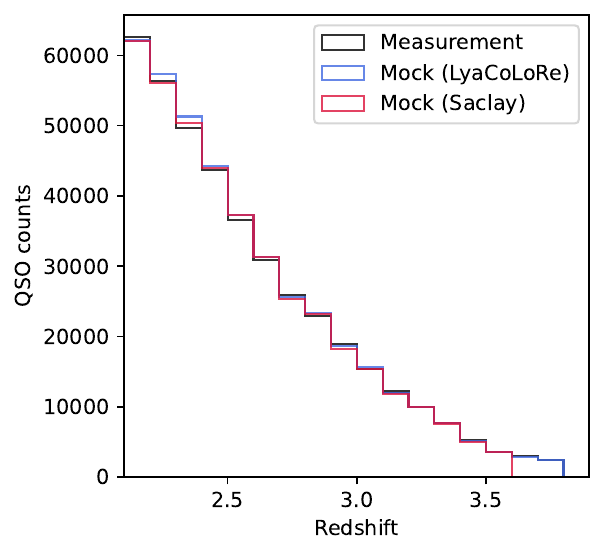}
\end{subfigure}
\caption{Left: Map of the locations of DESI observations for mock quasars in a DESI DR1 realization. The colours show the assigned effective exposure time. Right: Redshift distributions of one \lyacolore\ mock and one \saclay\ mock compared to DESI DR1 quasars over the same redshift range.}
\label{fig:mock_footprint}
\end{figure}

Using each of the 150 transmitted flux boxes described in \Cref{subsec:mock_boxes}, we generated a synthetic quasar spectra dataset following the same procedure as described in \citep{Herrera-Alcantar:2024}, with a modification on the method to mirror the observed footprint and object number density of the DESI DR1. We use the \texttt{desisim}\footnote{\url{https://github.com/desihub/desisim}} package to simulate both quasar populations and to produce synthetic spectra. The DESI Early Data Release plus two months of observations (EDR+M2) mocks presented in \citep{Herrera-Alcantar:2024} follow a method to emulate the footprint, object number density, and effective exposure time distribution which consists of dividing the observed footprint into HEALpix\footnote{\url{https://healpix.sourceforge.io}} pixels \cite{Gorski:2005} of \path{nside=16} and then downsampling the available mock quasars to match the density of observed data by HEALpix pixel. This method alters the shape of the quasar auto-correlation function because the exact number of quasars in each HEALpix pixel is related to the strength of the quasar clustering. Therefore, a downsampling factor computed from the ratio between the number of quasars from the simulated box and DESI DR1 over small patches will bias the resulting quasar auto-correlation. To address this issue, we have developed a new methodology that mirrors the spatial inhomogeneities introduced by the DESI survey strategy and does not alter the shape of any of the correlations.

The procedure is as follows. First, we randomly downsample the available targets in the input boxes so that the result follows the expected redshift distribution of the DESI survey and assign a random r-band magnitude to each target following the same procedure as described in Section 3 of \citep{Herrera-Alcantar:2024}. Then, we subdivide the sky into HEALpix pixels of \path{nside=1048}. For each pixel, we count the number of tiles on the DESI DR1 footprint that overlap in that region, we refer to this number as the number of passes ($N_{\text{passes}}$). In the observation strategy of the main DESI survey, the maximum number of tiles that can overlap in a given region is seven. The nominal effective exposure time for one DESI observation is 1000 seconds. However, through the DESI survey, Lyman-$\alpha$ quasars will be observed four times for a total effective exposure time of 4000 seconds~\cite{2023Schlafly:SurveyOps}. Once the number of passes as a function of position in the sky has been computed, we count the total number of observed ($N_{\text{data}}$) and mock ($N_{\text{mock}}$) quasars whose positions are in regions observed by $N_{\text{passes}}=1,2,...,7$ tiles. Finally, we randomly select the mock targets, whose spectra will be simulated, following the ratio $N_{\text{data}}/N_{\text{mock}}$ for each of the possible number of passes. At the same time, we compute a total exposure time probability distribution as a function of $N_{\text{passes}}$, based on the effective exposure time of the observed data quasars defined by $T_{\rm{eff}} = 12.15\ \rm{seconds} \times \text{TSNR}^2_{\rm{LRG}}$, where $\text{TSNR}_{\rm{LRG}}$ is the signal-to-noise ratio of the LRG template~\citep{Guy:2022wlv}. We use this probability distribution function to randomly assign an integer multiple of 1000 seconds effective exposure time to our mock quasars based on the number of passes corresponding to their position. The result is a preprocessed catalogue of the quasar targets we wish to simulate with exposure times and r-band magnitudes. Finally, we simulate non-linear peculiar velocities (the \textit{Fingers of God} effect) by adding random Gaussian velocities to our quasars, with a standard deviation of $150$ $\text{km}/\text{s}$. This only affects the \lya-QSO cross-correlation and is subdominant relative to the similar and larger impact of redshift errors \cite{Youles:2022}. The addition of redshift errors is discussed in \Cref{subsec:mock_spec} below.

The left panel of \Cref{fig:mock_footprint} shows the resulting footprint and exposure times when applying this procedure to a \lyacolore\ mock catalogue. The zones with a higher number of passes correspond to those with higher exposure times. Note that before applying the methodology to select the mock quasars, we restrict the redshift range of the observed data catalogue to match the range in the input boxes, $z<3.8$ for \lyacolore, and $z<3.6$ for \saclay. The comparison of the observed redshift distribution with the results from one realization of each type of mock is shown in the right panel of the same figure.

%%%%%%%%%%%%%%%%%%%%%%%%%%%%%%%%%%%%%%%%%%%%%%%%%%%%%%%%%%%%%%%%%%%%%%%%%%%%%%%%%%%%%%%%%%%%%%%%5

\subsection{Simulating quasar spectra}
\label{subsec:mock_spec}

Once we have a catalogue of quasars with associated \lya\ transmitted flux skewers, we want to turn these skewers into realistic DESI spectra. We use the \texttt{quickquasars}\footnote{\url{https://github.com/desihub/desisim/blob/main/py/desisim/scripts/quickquasars.py}} script from the \texttt{desisim}\footnote{\url{https://github.com/desihub/desisim}} package to generate our synthetic spectra. This script starts by taking the input transmitted flux boxes described in \cref{subsec:mock_boxes}, and post-processing them to introduce absorption features due to astrophysical sources (contaminants). Following \cite{Herrera-Alcantar:2024}, these include:
\begin{itemize}
    \item Damped \lya\ absorbers (DLAs) that are correlated with the density field (\Cref{subsec:mock_boxes}).
    \item Broad Absorption Lines (BALs) that are randomly added to $16\%$ of the targets \cite{Filbert2023}.
    \item Higher Lyman lines (up to Ly$\epsilon$) using rescaled versions of the \lya\ optical depth skewers based on the oscillator strengths of each transition.
    \item Metal absorbers using rescaled versions of the \lya\ optical depth skewers, with relative absorption strength coefficients ($C_m$) tuned through the process described in \Cref{subsec:metal_tune}. We model four metal absorption lines: SiII$(1190)$, SiII$(1193)$, SiIII$(1207)$, and SiII$(1260)$.
\end{itemize}
Section 2.3 of \citep{Herrera-Alcantar:2024} gives a detailed description of how astrophysical contaminants are included.

\begin{figure}
    \centering
    \includegraphics[width=1.0\textwidth,keepaspectratio]{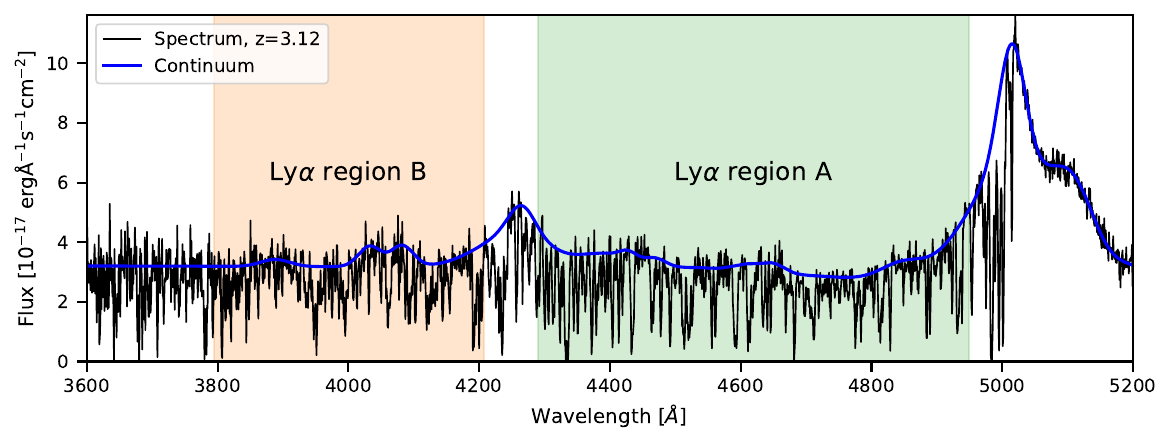}
    \caption{Example synthetic spectrum of a redshift $z=3.12$ quasar as obtained from \texttt{quickquasars}. The blue solid line shows the continuum template used to generate the spectrum prior to introducing instrumental noise. Coloured bands show the \lya\ regions A and B.}
    \label{fig:example_spectrum}
\end{figure}

Noiseless spectra are generated by multiplying the post-processed transmitted flux (now including contaminants) with templates for the quasars' unabsorbed spectrum, referred to as the continuum. We use the \texttt{SIMQSO} method in \texttt{quickquasars} to generate the continuum templates. At last, we use the \texttt{specsim}\footnote{\url{https://github.com/desihub/specsim}} package \cite{Kirkby:2016} to introduce instrumental noise to the spectra by emulating the specifications of the DESI spectrograph and nominal observational conditions set by the dark-time program of the main DESI survey. See Section 3.2.4 of \cite{Herrera-Alcantar:2024} for further details. \Cref{fig:example_spectrum} shows an example spectrum of a mock quasar at redshift $z=3.12$ produced with the aforementioned methodology.

We also generate in post-processing a contaminated quasar catalogue with random redshift errors following a Gaussian distribution with dispersion $\sigma_z=400\ \text{km/s}$. These have a different impact to the non-linear peculiar velocities added in \Cref{subsec:mock_qso} because they are added after the quasar continua are generated. This leads to the positions of emission lines being randomly shifted relative to the redshift used to generate the quasar continua. For our baseline mocks, we only use the redshift errors in the computation of the \lya-QSO cross-correlation, where they have a smoothing effect similar to non-linear peculiar velocities. However, following \cite{Youles:2022}, we also study their impact on the quasar continuum fitting process in \Cref{subsec:res_zerr}, where the random shifts in the position of the emission lines in the \lya\ forest region can introduce spurious correlations.

%%%%%%%%%%%%%%%%%%%%%%%%%%%%%%%%%%%%%%%%%%%%%%%%%%%%%%%%%%%%%%%%%%%%%%%%%%%%%%%%%%%%%%%%%%%%%%%%5

\subsection{Metal tuning}
\label{subsec:metal_tune}

To simulate the contamination by metal lines, we start with the \lya\ optical depth skewers and re-scale the optical depth by a scaling factor ($C_m$). These re-scaled skewers are then turned into transmitted flux skewers and shifted in wavelength such that they simulate absorption by the corresponding metal line. See \cite{Herrera-Alcantar:2024} for a detailed description of this process. One important factor to note is that the RSD effect is already present in the optical depth skewers used for this, which means metals in the mocks have an RSD signal similar to \lya, with RSD parameter $\beta\sim1.5$ \cite{Herrera-Alcantar:2024}. This is in contrast to real data, where metals are associated with galaxies, and therefore have a smaller RSD parameter, $\beta\sim0.5$. We will discuss the impact of this approximation later on in \Cref{subsec:model}.

For previous mock data sets, the re-scaling factor has been tuned such that the observed metal contamination in the resulting \lya\ correlation functions matches the metal contamination measured from eBOSS data \cite{duMasdesBourboux:2020,Herrera-Alcantar:2024}. The DESI DR1 \lyaf\ data set has significantly more quasar spectra than eBOSS, which means the measurements presented in \kplya are now the most precise \lyaf\ correlation function measurements. Therefore, we have used the DESI DR1 measurements to re-tune the values of $C_m$.

In order to match the simulated metal contamination in our mocks to the contamination observed in the data, we performed an iterative process that involves the following steps. First, we assume some values for the relative strength coefficients of the four metal lines we wish to simulate,\footnote{In practice we started from the values reported in \cite{Herrera-Alcantar:2024}.} and then create a set of 5 DESI DR1 mock data sets using the metal contamination set by those values. After that, we measured the \lya\ correlation functions (all four correlations presented in \Cref{subsec:corr}), and jointly fit these correlations using the model described in \Cref{subsec:model} to measure the linear bias parameters for the four metal lines. \cite{Herrera-Alcantar:2024} found a linear relation between the measured bias parameters and the relative strengths $C_m$, which means we can use the ratio between the biases measured in mocks versus data to compute the next $C_m$ estimates. We found that we only need to perform this process once or twice to be able to reproduce the measured metal biases from \kplya (i.e. it quickly converges).

\begin{table}[]
    \centering
    \begin{tabular}{c|c|c|c|c}
        % \hline
         & & \multicolumn{2}{|c|}{Relative strength ($C_m\times 10^3$) } &  \\
        % \hline
        Metal line & $\lambda_m$ [\AA] & \lyacolore\ & \saclay\ & $\Delta r_{||}$ [\hMpc] \\
        \hline%\hline
        SiIII & 1207 & 3.5 & 1.6 & -21 \\
        % \hline
        SiII & 1190 & 1.4 & 0.68 & -53 \\
        % \hline
        SiII & 1193 & 0.7 & 0.53 & -60 \\
        % \hline
        SiII & 1260 & 1.3 & 0.57 & 103 \\
        % \hline
    \end{tabular}
    \caption{The four metal lines we use to simulate metal absorption contaminating the \lyaf\ region. The contamination from these metal lines is based on a re-scaling of \lya\ optical depth skewers. We show the relative strength coefficients ($C_m$) used to perform this re-scaling for both \lyacolore\ and \saclay\ mocks. We also show the effective separation in comoving coordinates between each metal line and the \lya\ line.}
    \label{tab:metal_strengths}
\end{table}

We show the resulting values for the relative strength coefficients $C_m$ of the four silicon lines we simulate in \Cref{tab:metal_strengths}. The tuning process was performed independently for \lyacolore\ and \saclay\ mocks, because we do not expect the metal contamination to be the same in both types of mocks. Indeed, the values of $C_m$ we obtain for the two types of mocks are fairly different. This can be explained by the fact that the fit of the metal biases is driven by the very small-scale line-of-sight cross-correlation between metal absorption and either \lya\ or quasars (i.e. the metal peaks present in \lya\ correlation functions). As described in \Cref{subsec:mock_boxes}, the main differences between \saclay\ and \lyacolore\ mocks are due to the strength of the small-scale quasar clustering (which affects metal peaks in the \lya-QSO cross-correlation) and the RSD signal (which affects line-of-sight correlations). \Cref{tab:metal_strengths} also contains the effective difference in co-moving coordinates between the \lya\ line and each metal transition. This difference is given by the separation along the line-of-sight at which we see the \lya-Metal cross-correlation peak in our measured correlation functions (see \Cref{subsec:model}). Note that these values are slightly different for DESI compared to BOSS and eBOSS because they rely on the redshift distribution of our pixel pairs \cite{duMasdesBourboux:2020,DESI2024.IV.KP6}.

%%%%%%%%%%%%%%%%%%%%%%%%%%%%%%%%%%%%%%%%%%%%%%%%%%%%%%%%%%%%%%%%%%%%%%%%%%%%%%%%%%%%%%%%%%%%%%%%5

\section{Analysis}
\label{sec:analysis}

As the main goal of this work is to validate the DESI \lyaf\ BAO measurement, our analysis process follows closely the analysis done on DESI DR1 data (\kplya). The first two parts of the analysis, involve computing the \lya\ flux overdensity ($\delta$) field (\Cref{subsec:deltas}) and the 3D correlation functions (\Cref{subsec:corr}). To compute these, we use the publicly available \picca\footnote{\url{https://github.com/igmhub/picca}} package. The algorithm behind \picca\ has been described in detail in \cite{duMasdesBourboux:2020} and \cite{Ramirez-Perez:2024}. Therefore, we only give a brief overview here, focusing on the parts that are most relevant to our analysis.

One of the main improvements in the DESI DR1 analysis is that we now take into account the cross-covariance between the different correlation functions. This means we need to compute a covariance matrix that covers all four correlation functions. We describe the process for computing this larger covariance matrix in \Cref{subsec:cov_mat}.

The final step of our analysis involves building a model for the correlation functions we have computed, and fitting for the BAO signal. This is achieved using the publicly available \vega\footnote{\url{https://github.com/andreicuceu/vega}} package. We give a detailed description of this model in \Cref{subsec:model}. While the analysis process up to this point is the same as the one used for the real DESI data, our model is slightly different than the one used in \kplya. These small differences are described in detail below, and we also discuss their impact on BAO measurements in \Cref{sec:discussion}.

\subsection{The Ly$\alpha$ flux overdensity field}
\label{subsec:deltas}

Before measuring the flux overdensity field, we first mask BALs and DLAs present in the spectra. We use the true BAL and DLA catalogues because running the BAL and DLA finders on the entire set of mocks is not computationally feasible. However, we have tested running them on one individual mock and we found it has negligible impact on the BAO measurement from that mock. For studies on the performance of the DLA finder in the context of DESI mocks see \cite{Wang:2022}, and for a detailed description of the BAL and DLA masking process see \cite{Ramirez-Perez:2024}. Following \cite{Ennesser:2022}, we keep all BAL quasars and mask their absorption features. For DLAs, we follow \cite{Ramirez-Perez:2024} and mask all DLAs with column densities $\log N_{\mathrm{H}\textsc{i}}>20.3$.\footnote{Note that \kplya imposes a signal-to-noise selection to ensure the purity of the DLA catalogue. As we use the true DLA catalogue here, we do not use this selection.} The mask is applied to the region where the DLA leads to more than $20\%$ absorption, while the rest of the DLA wings are corrected using a Voigt profile \cite{Noterdaeme:2012}.

The flux overdensity field in the spectrum of a quasar $q$ at observed wavelength $\lambda$ is given by:
\begin{equation}
    \delta_q(\lambda) = \frac{f_q(\lambda)}{\overline{F}(\lambda)C_q(\lambda)}-1,
\end{equation}
where $f_q$ is the measured flux, $\overline{F}$ is the global mean \lya\ flux, and $C_q$ is the quasar continuum. In general, we do not know the true quasar continuum, so we fit it along with $\overline{F}$ directly from the data \cite{duMasdesBourboux:2020,Ramirez-Perez:2024}. This involves expressing the product $\overline{F}(\lambda)C_q(\lambda)$ as:
\begin{equation}
\overline{F}(\lambda)C_q(\lambda)=\overline{C}(\lambda_{\rm RF})\left(a_q + b_q\frac{\Lambda - \Lambda_{min}}{\Lambda_{max} - \Lambda_{min}}\right),
\end{equation}
where $\overline{C}(\lambda_{\rm RF})$ is a universal function of rest-frame wavelength ($\lambda_{\rm RF}$) and $\Lambda\equiv\log\lambda$. The parameters $a_q$ and $b_q$ are the amplitude and slope that we fit individually for each quasar spectrum in order to account for quasar spectral diversity. This fit also requires an estimate of the total flux variance $\sigma^2_q(\lambda)$. Following \cite{Ramirez-Perez:2024}, this is given by:
\begin{equation}
    \sigma_q^2(\lambda) = \eta_{\rm pip}(\lambda) \sigma_{\rm pip,q}^2(\lambda) + \sigma_{\rm LSS}^2(\lambda) [\overline{F}(\lambda)C_q(\lambda)]^2,
\end{equation}
where $\sigma_{\rm pip,q}$ is usually the flux variance estimated by the DESI pipeline, but in our case it is the variance of the simulated noise in our synthetic spectra. $\eta_{\rm pip}(\lambda)$ is a correction factor meant to account for inaccuracies in the variance estimate, and $\sigma_{\rm LSS}(\lambda)$ is the intrinsic large-scale structure (LSS) variance.

The process of continuum fitting involves an iteration that starts with an estimate of the global quantities $\overline{C}(\lambda_{\rm RF})$, $\eta_{\rm pip}(\lambda)$, and $\sigma_{\rm LSS}(\lambda)$. We then fit the $a_q$ and $b_q$ parameters for each spectrum and measure the $\delta_q(\lambda)$. After that, we measure the variance of this field and fit for the $\eta_{\rm pip}(\lambda)$ and $\sigma_{\rm LSS}(\lambda)$ functions. Finally, we measure the global mean continuum $\overline{C}(\lambda_{\rm RF})$, and then repeat this iterative process until convergence. In practice, 5 steps are enough to achieve convergence \cite{duMasdesBourboux:2020}. For a detailed description of this process see \cite{Ramirez-Perez:2024}.

We measure the \lya\ flux overdensity field in two distinct regions, which we refer to as regions A and B. Region A is located between the \lya\ and \lyb\ peaks in the rest-frame wavelength interval $1040-1205$ \AA. Region B is located between the \lyb\ peak and the Lyman limit, in the interval $920-1020$ \AA. Note that even though region B also contains \lyb\ and higher order absorption lines,\footnote{The presence of these higher order absorption lines only introduces extra noise due to the large comoving separation relative to the \lya\ absorption.} we only work with the \lya\ flux here. Therefore, we will use the symbol Ly$\alpha$(A) for \lya\ region A, and Ly$\alpha$(B) for \lya\ region B. The continuum fitting process is performed separately for the two regions. For a comparison of the performance of this continuum fitting method in our mocks versus the real data, see  Section 4.2 of \cite{Herrera-Alcantar:2024}.

The continuum fitting process also has an unintended effect that has a large impact on the measured correlation functions. This arises due to the amplitude and slope parameters that we fit to each quasar spectrum. While the purpose of these parameters is to account for quasar spectral diversity, they also capture some large-scale structure information. In particular, this model will also fit any large-scale mode of the size of the forest and larger that is present in the data. This biases the measured $\delta$ towards zero, and results in a distortion of the measured correlation functions. Following \cite{Bautista:2017}, we account for this distortion by building projection matrices $\eta_{ij}$, such that:
\begin{equation}
    \sum_j \eta_{ij}\delta^m(\lambda_j) = \sum_j \eta_{ij}\delta^t(\lambda_j),
    \label{eq:projection}
\end{equation}
where $\delta^m$ is the measured flux overdensity after continuum fitting, $\delta^t$ is the true flux overdensity. The equality in \Cref{eq:projection} is not exact on a per-forest basis due to noise, but the formalism is built on the assumption that the two sides converge to equality when averaging over a large enough sample of forests (as we do when we compute correlation functions).
The projection matrices are given by:
\begin{equation}
    \eta_{ij} = \delta_{ij}^K - \frac{w_j}{\sum_k w_k} - \frac{w_j\kappa_i\kappa_j}{\sum_k w_k \kappa_k^2},
\end{equation}
where $\delta_{ij}^K$ is the Kronecker delta, $\kappa_k=\log \lambda_k - \overline{\log \lambda_q}$, and the weights $w_i$ are described below. For a detailed description of the projection and the assumptions behind it see \cite{Bautista:2017} and \cite{Perez-Rafols:2018}. Using this formalism, we project the measured flux overdensity field using the left-hand side of \Cref{eq:projection}. However, we model correlation functions (not the $\delta$ field), so the right-hand side is instead propagated into our correlation function model. We describe this process in \Cref{subsec:model}.

The weights used to build projection matrices are the same weights we use to compute correlation functions, and are given by:
\begin{equation}
    w(\lambda_i) = \frac{[(1+z_i)/(1+2.25)]^{\gamma_{{\rm Ly}\alpha}-1}}{ \eta_{\rm LSS} \sigma_{\rm LSS}^2(\lambda) + \eta_{pip}(\lambda) \Tilde{\sigma}_{pip,q}^2(\lambda)},
    \label{eq:lya_weights}
\end{equation}
where $\Tilde{\sigma}_{pip,q}^2 = \sigma_{pip,q}^2 / [\overline{F}(\lambda)C_q(\lambda)]^2$, and we take into account the redshift evolution of the \lya\ bias using $\gamma_{{\rm Ly}\alpha}=2.9$ based on \cite{McDonald:2006}. The $\eta_{\rm LSS}$ term is an ad-hoc correction factor that modulates the relative importance of instrumental noise versus intrinsic \lya\ fluctuations. This was introduced following the study by \cite{Ramirez-Perez:2024} to minimize the uncertainties in the correlation function estimates. For our dataset, \cite{Ramirez-Perez:2024} found $\eta_{\rm LSS}=7.5$ to be the optimal value.

%%%%%%%%%%%%%%%%%%%%%%%%%%%%%%%%%%%%%%%%%%%%%%%%%%%%%%%%%%%%%%%%%%%%%%%%%%%%%%%%%%%%%%%%%%%%%%%%5

\subsection{Estimating correlation functions}
\label{subsec:corr}

\begin{figure*}
\includegraphics[width=1.0\linewidth]{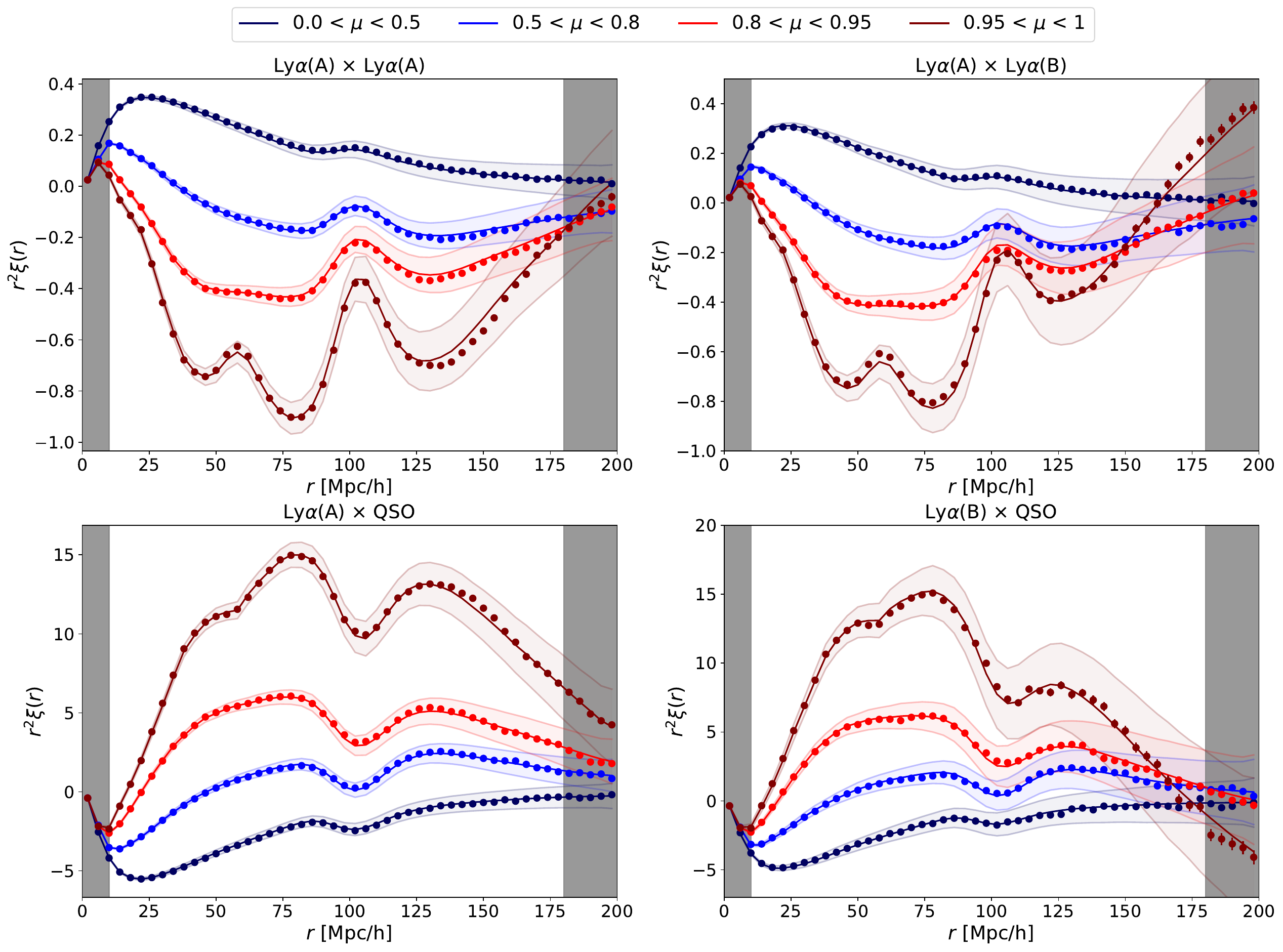}
\caption{Stacked correlation functions from a set of 100 DESI DR1 \lyacolore\ mocks compressed into $\mu=r_{||}/r$ wedges and shown as a function of isotropic separation $r$ (points with error-bars). The four panels each show one of the four correlation functions we compute, with the two auto-correlations on the top row, and the two cross-correlations on the bottom row. The shaded regions indicate the DESI DR1 uncertainties, and the lines indicate the best-fit model described in \Cref{subsec:model}.}
\label{fig:colore_stack}
\end{figure*}

\begin{figure*}
\includegraphics[width=1.0\linewidth]{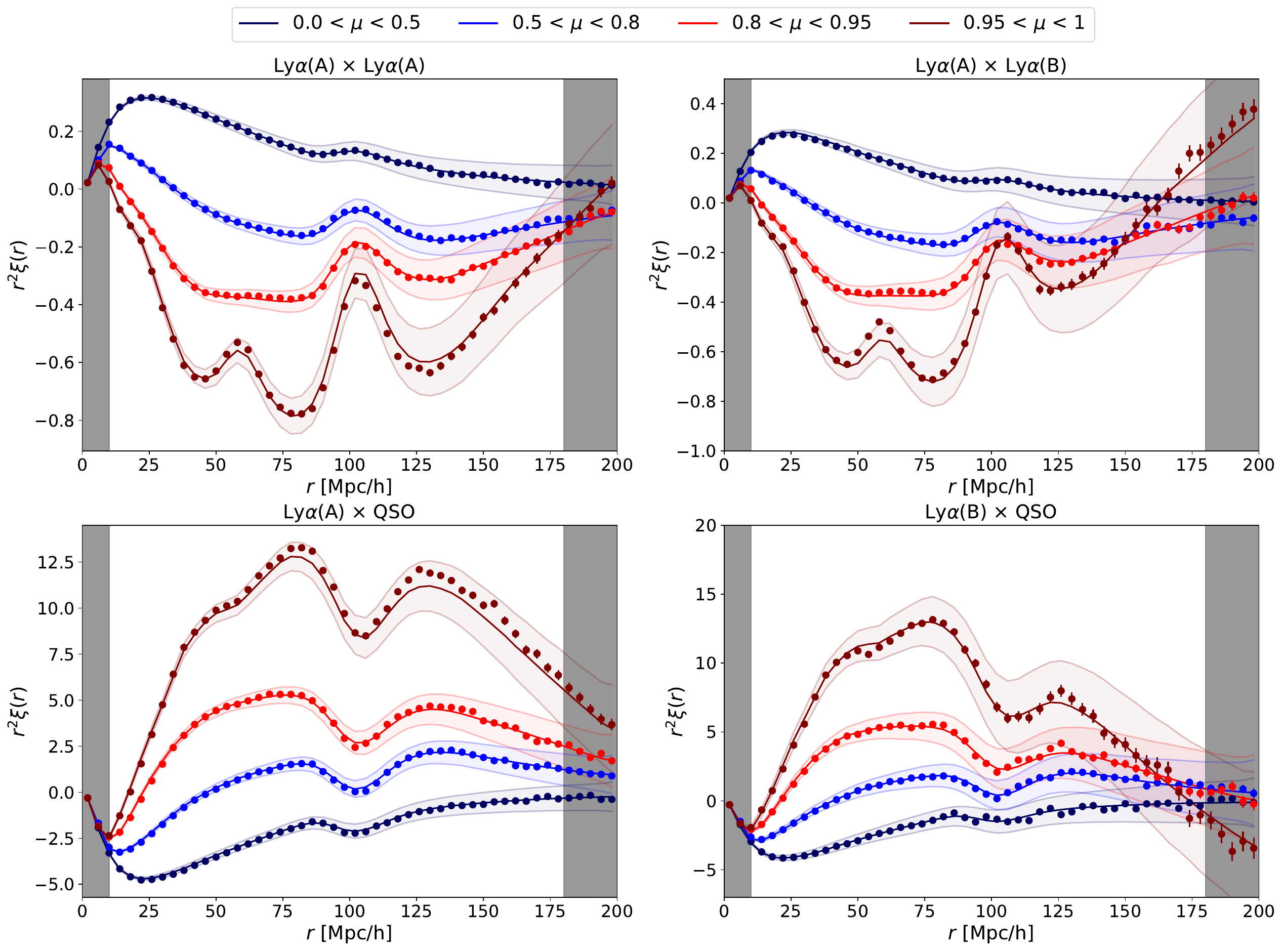}
\caption{Similar to \Cref{fig:colore_stack}, but showing the stack of correlation functions measured from 50 DESI DR1 \saclay\ mocks.}
\label{fig:saclay_stack}
\end{figure*}

For the DESI high redshift BAO measurement, we use both the \lyaf\ and quasars as tracers of large-scale structure. As discussed above, we measure the \lya\ flux overdensity field in two separate regions of our spectra. Therefore, we have six different two-point functions that can be computed from these three datasets (\lya(A), \lya(B), and QSOs). Following \cite{duMasdesBourboux:2020}, we focus on a subset of four correlations. These include two auto-correlations of \lya\ flux, \lyalyalyalya\ and \lyalyalyalyb, and two cross-correlations between \lya\ flux and quasars, \lyalyaqso\ and \lyalybqso.

We compute correlation functions on a grid in comoving separation along ($r_{||}$) and across ($r_\bot$) the line-of-sight. These are computed from the measured angles $\theta$ and redshifts $z$ using a fiducial cosmology. For two pixels, $i$ and $j$, separated by $\Delta\theta$, the comoving separations are given by \cite{deSainteAgathe:2019}:
\begin{align}
    r_{||} &= [D_\mathrm{c}(z_i) - D_\mathrm{c}(z_j)] \cos{\frac{\Delta \theta}{2}}, \\
    r_\bot &= [D_\mathrm{M}(z_i) + D_\mathrm{M}(z_j)] \sin{\frac{\Delta \theta}{2}},
\end{align}
where $D_\mathrm{c}$ and $D_\mathrm{M}$ are the radial and transverse comoving distances. As mentioned in \Cref{subsec:mock_boxes}, we use the Planck 2015 results \cite{Planck:2016} as our fiducial cosmology. Besides $r_{||},r_\bot$, we will use the $r,\mu$ parametrization when presenting correlation functions, where $r^2=r_{||}^2+r_\bot^2$ and $\mu=r_{||}/r$. We compute correlation functions in bins of $4$ \hMpc\ in both $r_{||}$ and $r_\bot$. For auto-correlations, we use a grid between $0$ and $200$ \hMpc\ for both coordinates, resulting in a $50\times50$ grid. On the other hand, for cross-correlations we distinguish between pixels in front of (negative $r_{||}$) and behind quasars (positive $r_{||}$), so $r_{||}$ takes values between $-200$ and $200$ \hMpc, resulting in a $100\times50$ grid.

To measure correlation functions, we follow previous \lyaf\ BAO analyses \cite{Bautista:2017,duMasdesBourboux:2020,Gordon:2023} and use a weighted pair-counting algorithm. The \lya\ flux auto-correlation and its cross-correlation with quasars are given by:
\begin{equation}
    \xi_M = \frac{\sum_{i,j \in M} w_i w_j \delta_i \delta_j}{\sum_{i,j \in M} w_i w_j},
\end{equation}
where $M$ defines a bin in comoving coordinates, and $\delta=1$ for quasars. The sums run over pixel-pixel pairs for the auto-correlation and over pixel-QSO pairs for the cross-correlation. The weights for the \lyaf\ are given by \Cref{eq:lya_weights}, while for quasars they are given by:
\begin{equation}
    w_Q = [(1+z_Q)/(1+2.25)]^{\gamma_Q-1},
\end{equation}
where $\gamma_Q=1.44$ based on measurements by \cite{dumasdesbourboux:2019}.

For our validation study, we also compute the mean and covariance of correlation functions from sets of many mocks. We refer to these as stacked correlation functions. This gives us measurements of the correlation function with negligible statistical uncertainties. Therefore, they can be used to test our model and BAO measurement with much greater statistical precision than an individual mock would allow (\Cref{subsec:res_stack}). We show the stacked correlation functions from the set of 100 \lyacolore\ mocks in \Cref{fig:colore_stack}, and from the 50 \saclay\ mocks in \Cref{fig:saclay_stack}. For a comparison of correlations functions measured from \lyacolore\ and \saclay\ mocks, as well as DESI data, see \cite{Herrera-Alcantar:2024}.

%%%%%%%%%%%%%%%%%%%%%%%%%%%%%%%%%%%%%%%%%%%%%%%%%%%%%%%%%%%%%%%%%%%%%%%%%%%%%%%%%%%%%%%%%%%%%%%%5

\subsection{Covariance matrix}
\label{subsec:cov_mat}

The biggest change between the DESI 2024 \lya\ BAO analysis and previous analyses on DESI EDR and SDSS data is how we compute the covariance matrix. The change is that we now compute one covariance matrix for all four correlation functions, which means we also take into account the cross-covariances between the individual correlations. In previous datasets these were found to be negligible \cite[e.g.][]{duMasdesBourboux:2020}. However, that is not the case for DESI DR1, where ignoring these cross-covariances leads to a $\sim10\%$ change in BAO uncertainties. The analysis that led to this decision is described in \kplya.

To compute the $15000 \times 15000$ covariance matrix we follow the same approach used in previous analyses \cite[e.g.][]{Bautista:2017,duMasdesBourboux:2017,duMasdesBourboux:2020}. We first compute correlation functions independently in each HEALPix pixel. For the DESI DR1 dataset, there are $1028$ pixels (\texttt{nside} $=16$), each covering a roughly $250\times250 \;(h^{-1}\mathrm{Mpc})^2$ patch at $z_\mathrm{eff}=2.33$. We then compute a noisy estimate of the covariance matrix from this sample of correlation function measurements:
\begin{equation}
    C_{MN} = \frac{1}{W_M W_N} \sum_s W_M^s W_N^s [\xi_M^s\xi_N^s - \xi_M\xi_N],
\end{equation}
where $s$ is a sub-sample, $W_M^s = \sum_{i\in M,s} w_i$, and $W_M = \sum_s W_M^s$. Finally, the noisy estimate of the covariance is smoothed at the level of the correlation matrix given by:
\begin{equation}
    \text{Corr}_{MN}\equiv C_{MN} / (C_{MM}C_{NN})^{1/2},
    \label{eq:corr_matrix}
\end{equation}
where $C_{MM}$ and $C_{NN}$ are the variances in bins $M$ and $N$ respectively. The smoothing is done by replacing non-diagonal elements of the correlation matrix which correspond to the same differences $|r_{||}(M)-r_{||}(N)|$ and $|r_\bot(M)-r_\bot(N)|$ with their average. This method has proven effective for obtaining estimates of the covariance matrices of individual correlations in the past \cite[see][]{Delubac:2015,duMasdesBourboux:2020}. However, for DESI DR1 we employ it to obtain an estimate of the much larger covariance matrix of all four correlations.

\begin{figure}
\centering
\includegraphics[width=0.8\linewidth]{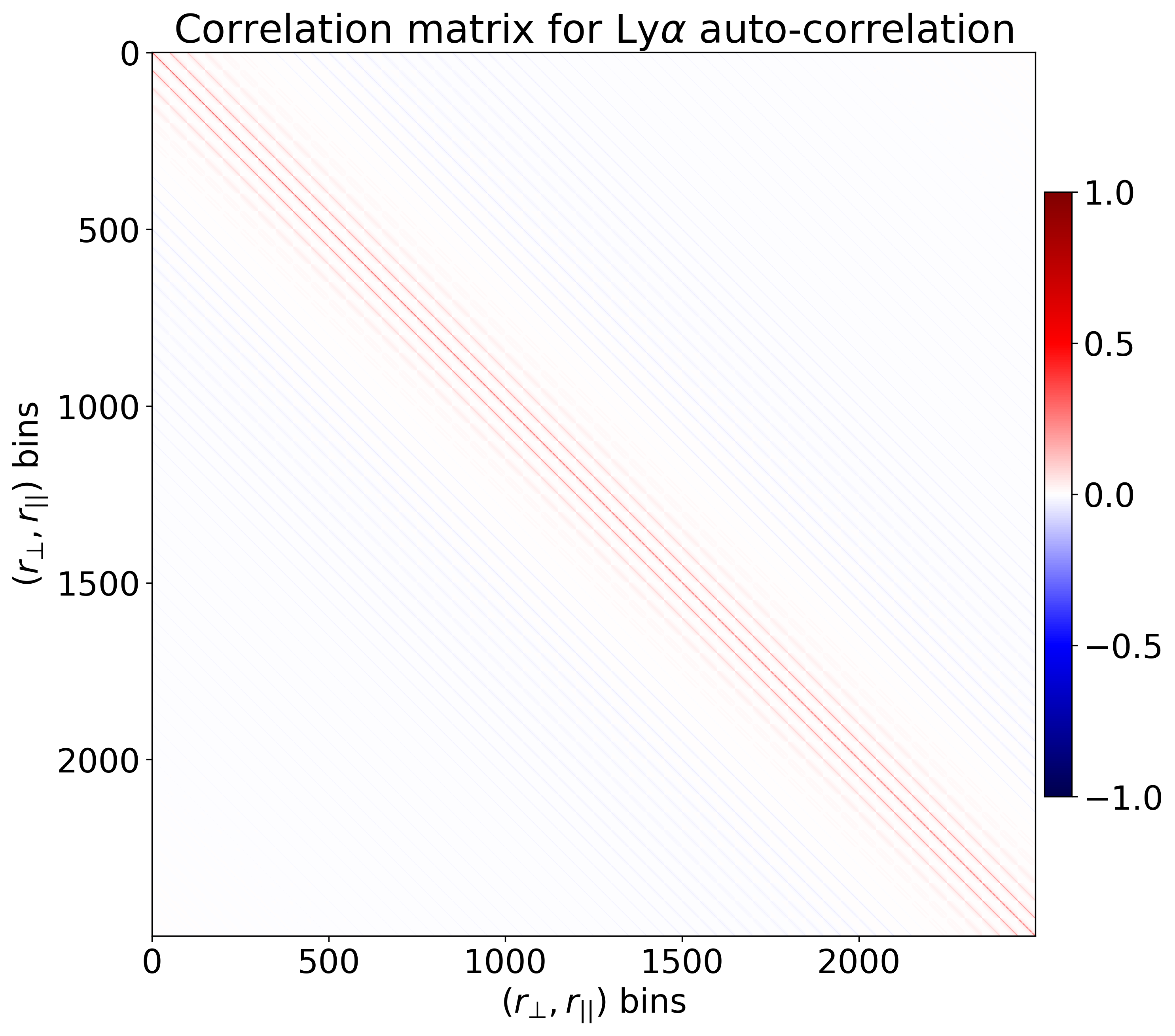}
\caption{Normalized smoothed covariance matrix (correlation matrix) for the \lya\ auto-correlation function computed from the stack of 100 \lyacolore\ mocks. This is part of the larger correlation matrix of all four \lya\ correlations. The off-diagonal lines visible in this figure are the correlations between line-of-sight bins for the same $r_\bot$. The most important correlations are between neighbouring $r_{||}$ bins (i.e. $\Delta r_{||}=4$\hMpc) which are $\sim0.3$ in magnitude (first off-diagonal lines visible here), with correlations at larger $\Delta r_{||}$ rapidly decaying to below $|0.1|$.}
\label{fig:correlation_matrix_auto}
\end{figure}

Using this procedure we compute an individual smoothed covariance matrix for each of the 150 mocks. We use separate covariance matrices for each mock in order to mimic the analysis of the DESI DR1 data (\kplya). For the stacked correlation functions, we first gather the correlation function samples from each mock and then compute the covariance matrix using the same method. This means we use much larger sets of samples for the covariance matrices of the stacked correlations ($\sim100$k for \lyacolore, and $\sim50$k for \saclay). We will use these stacked covariance matrices to test our method for estimating the full covariance of individual mocks in \Cref{subsec:res_cov}.

The covariance matrices for 2 out of the 150 mocks are not positive semi-definite even after smoothing. In order to use them for our analysis, we combine the correlation matrix from the stack of mocks with the variance estimates of each of the two mocks.\footnote{This is the same approach we use to test our covariance estimates in \Cref{subsec:res_cov}. These covariance matrices are given by \Cref{eq:corr_mat_test}.} Note that this only affects our studies of the population of mock results, and not the results from the stacked correlation functions. Furthermore, we have also performed the entire analysis with the two mocks discarded, and it did not significantly impact any of our results or conclusions.

\begin{figure}
\centering
\includegraphics[width=0.8\linewidth]{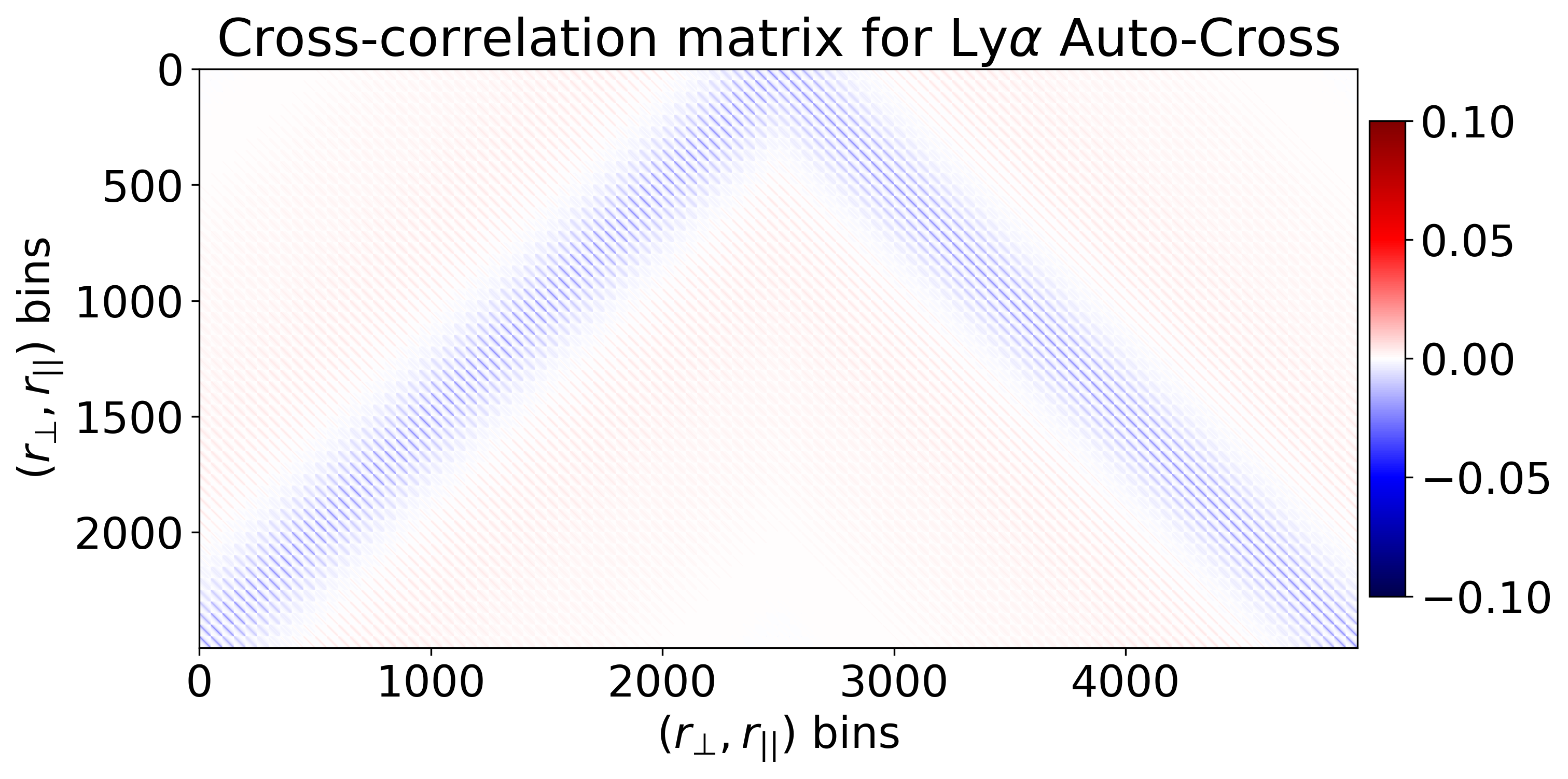}
\caption{Smoothed cross-correlation matrix between the \lya\ auto-correlation and the \lya-QSO cross-correlation from the stack of 100 \lyacolore\ mocks. Note that the range of the colour scale is 10 times smaller than \Cref{fig:correlation_matrix_auto}. Unlike previous \lya\ BAO analyses, with DESI DR1 this cross-correlation needs to be taken into account, as it leads to $\sim10\%$ changes in the BAO uncertainties. The structure can be understood when considering that the \lya-QSO cross-correlation is computed on a grid from $-200$\hMpc\ to $200$\hMpc, while the \lya\ auto-correlation is symmetric along the line-of-sight, and therefore computed on a grid from $0$\hMpc\ to $200$\hMpc. The most important correlations are again between neighbouring bins along the line-of-sight (i.e. $\Delta r_{||}=4$\hMpc, $\Delta r_\bot=0$\hMpc).}
\label{fig:correlation_matrix_auto_cross}
\end{figure}

The smoothed correlation matrix of the \lya\ auto-correlation function from the stack of 100 \lyacolore\ mocks is shown in \Cref{fig:correlation_matrix_auto}. For visualization purposes, we only show a subset of the larger correlation matrix. We also show the cross-correlation matrix between \lyalyalyalya\ and \lyalyaqso\ in \Cref{fig:correlation_matrix_auto_cross}. The features that dominate the covariance matrix are the off-diagonal lines at regular intervals of 50 bins. These represent the correlations between line-of-sight bins at the same transverse separation (i.e. $\Delta r_{||}>0$\hMpc, $\Delta r_\bot=0$\hMpc). The most important of these are correlations between neighbouring line-of-sight bins ($\Delta r_{||}=4$\hMpc), which are $\sim0.3$ in magnitude (for the auto-covariance). Correlations at larger separations ($\Delta r_{||}>4$\hMpc) rapidly decay below $|0.1|$. We discuss and provide other illustrations of our correlation matrix in \Cref{subsec:res_cov}.

%%%%%%%%%%%%%%%%%%%%%%%%%%%%%%%%%%%%%%%%%%%%%%%%%%%%%%%%%%%%%%%%%%%%%%%%%%%%%%%%%%%%%%%%%%%%%%%%5

\subsection{Modelling correlation functions}
\label{subsec:model}

Our model for the \lyaf\ correlation functions is based on a template approach that was first introduced in \cite{Kirkby:2013}. We start with an isotropic linear matter power spectrum $P_\mathrm{fid}$, that is split into a peak (or wiggles) component $P_\mathrm{fid}^\mathrm{p}$, and a smooth (or no-wiggles) component $P_\mathrm{fid}^\mathrm{s}$. This represents our template. The modelling process involves adding the Kaiser term \cite{Kaiser:1987}, as well as models for non-linearities, some contaminants, and various other effects we need to account for. The now anisotropic power spectrum model is then transformed into a correlation function model and interpolated on a grid in $r_{||}$ and $r_\bot$. Finally, we add the effect of metal contamination and distortion due to continuum fitting. The entire process is performed separately for the smooth and peak components of the template, and the final step involves combining them. The BAO feature is measured by allowing the coordinates ($r_{||},r_\bot$) of the peak component to vary using two scale parameters (\apar,\atrans) which are described below. We use the \vega\ package to model and fit our correlation functions.

The isotropic linear matter power spectrum $P_\mathrm{fid}$ is computed using \texttt{CAMB} \cite{Lewis:1999} and the same fiducial cosmology above \cite{Planck:2016}. The decomposition into peak and smooth components is performed using the algorithm described in \cite{Kirkby:2013}. The anisotropic model cross-spectrum is given by:
\begin{multline}
P_{A \times B}(k,\mu_k,z) = b_A  b_B (1 + \beta_A\mu_k^2) (1 + \beta_B\mu_k^2)
\;G(k,\mu_k) F_\mathrm{SM}(k, \mu_k) F_\mathrm{NL}(k, \mu_k) P_\mathrm{fid}(k), 
\label{eq:power_spec}
\end{multline}
where ($A$, $B$) are either (\lya, \lya) for auto-correlations, or (\lya, QSO) for cross-correlations. $b_X$ and $\beta_X$ are the linear bias and RSD parameters (for $X$ either \lya\ or QSO), and $\mu_k=k_{||}/k$ with isotropic wavenumber $k$, and line-of-sight wavenumber $k_{||}$. $G(k,\mu_k)$ models the binning of the correlation function and is given by: $G(k,\mu_k) = \mathrm{sinc}(k_{||}R_{||}/2)\;\mathrm{sinc}(k_\bot R_\bot/2)$, with bin sizes $R_{||}=R_\bot=4$ \hMpc. $F_\mathrm{SM}$ and $F_\mathrm{NL}$ are the smoothing and redshift error models respectively, and are described below.

The \lya\ $\delta$ field also contains absorption from unmasked HCDs that are too small to be detected. As these HCDs also trace the underlying large-scale structure, their auto- and cross-correlations with \lya\ and QSOs need to be modelled. Following \cite{Font-Ribera:2012b}, these can be added to our model by simply treating the bias and RSD parameters associated with the \lyaf\ as effective parameters ($b'_{\mathrm{Ly}\alpha},\beta'_{\mathrm{Ly}\alpha}$) that include both \lya\ and HCDs.\footnote{\cite{Font-Ribera:2012b} found that some higher-order functions can also have a small but detectable impact. However, similar to previous analyses, we ignore these and only model 2-point functions.} These are given by:
\begin{align}
    b'_{\mathrm{Ly}\alpha} &= b_{\mathrm{Ly}\alpha} + b_\mathrm{HCD} F_\mathrm{HCD}(k_{||}), \label{eq:hcd_bias} \\
    b'_{\mathrm{Ly}\alpha}\beta'_{\mathrm{Ly}\alpha} &= b_{\mathrm{Ly}\alpha}\beta_{\mathrm{Ly}\alpha} + b_\mathrm{HCD} \beta_\mathrm{HCD} F_\mathrm{HCD}(k_{||}), \label{eq:hcd_beta}
\end{align}
where parameters with subscript \lya\ are associated with the IGM, and parameters with subscript HCD are associated with high column-density absorbers. The function $F_\mathrm{HCD}(k_{||})$ depends on the column density distribution function of the HCDs present in our data \cite{McQuinn:2011,Font-Ribera:2012b}. However, measurements of this function in the $N_{\mathrm{H}\textsc{i}}$ range of interest ($\log N_{\mathrm{H}\textsc{i}}<20.3$) are very limited \cite{Noterdaeme:2009,Prochaska:2010,Noterdaeme:2012,Zafar:2013}. Therefore, we use the approximate form $F_\mathrm{HCD} = \exp(-L_\mathrm{HCD}k_{||})$, where $L_\mathrm{HCD}$ can be interpreted as the typical length scale of unmasked HCDs \cite{deSainteAgathe:2019}. Following \cite{Cuceu:2020}, and in line with \kplya, we treat $L_\mathrm{HCD}$ as a free parameter that we marginalize over. For a comparison of the different HCD models, see Appendix A of \kplya.

In our mocks, the $F_\mathrm{NL}(k, \mu_k)$ term only applies to the \lya-QSO cross-correlation and models the statistical quasar redshift errors and quasar non-linear velocities. For the analysis on data, \kplya test both a Lorentzian and a Gaussian smoothing, following \cite{Percival:2009}. These are given by:
\begin{align}
    F^2_\mathrm{NL,Lorentz} &= [1 + (k_{||}\sigma_z)^2]^{-1}, \\
    F^2_\mathrm{NL,Gauss} &= \exp \bigg[-\frac{1}{2}(k_{||}\sigma_z)^2 \bigg],
\end{align}
where $\sigma_z$ is a free parameter. As our redshift errors were injected using a Gaussian distribution, we use that form for the $F_\mathrm{NL}(k, \mu_k)$ term. However, we also tested the Lorentzian distribution and found  that the choice between these two functional forms does not have an impact on BAO measurements. Note that for the analysis on data, \kplya also used a non-linear term for the small scales in the auto-correlation based on \cite{Arinyo:2015}. However, we do not use it because our mocks are based on a Gaussian field and therefore do not have the same small-scale non-linearities present in the real data. We have tested adding this model and we found it has no impact on BAO measurements.

The input log-normal simulations have a grid cell size of $\sim 2.4$ \hMpc, which results in extra smoothing of the measured correlations functions. Following \cite{Farr:2020}, we add Gaussian anisotropic smoothing to account for this effect, represented by the $F_\mathrm{SM}$ term in \Cref{eq:power_spec}. This model has two smoothing scale parameters $(\sigma_{||}, \sigma_\bot)$, which we fit and marginalize over. This is the only component of our model that is present in the analysis on mocks, and absent from the analysis on data.

The next step in the modelling process involves transforming the anisotropic model power spectrum into a model correlation function. This is done by first performing a multipole decomposition up to $\ell=6$, followed by a Hankel transform,\footnote{Using the FFTLog algorithm \cite{Hamilton:2000} with the \texttt{mcfit} package \url{https://github.com/eelregit/mcfit}.} and finally computing the two-dimensional correlation function from the correlation multipoles. We have also tested including multiples with $\ell>6$, and found no impact on BAO measurements. We interpolate the model onto a grid with $2$\hMpc\ bins in $r_{||}$ and $r_\bot$. This is an improvement over previous analyses which used $4$\hMpc\ bins and allows us to build a more precise model for the correlation functions.

Besides \lya\ flux, we also model the contamination due to metal absorption. This involves computing correlation function models for all \lya-Metal and Metal-Metal cross-correlations using the same framework as for the main \lya\ correlations. Each metal line has a separate linear bias and RSD parameters $(b,\beta)$, but we neglect HCD effects for the \lya\ part of these correlations. Following previous analyses, we fix the metal RSD parameters to 0.5 \cite{Bautista:2017,duMasdesBourboux:2017,duMasdesBourboux:2020}. However, given that the metals in our mocks are added by re-scaling the \lya\ flux field, their RSD parameters are likely closer to the \lya\ RSD parameter ($\sim1.5$). We have tested that this has a negligible impact on our results. When computing correlation functions we assume all pixels are caused by \lya\ absorption. As some of the absorption is caused by metal lines, we assign these pixel pairs to the wrong correlation function bins. Following \cite{Bautista:2017,duMasdesBourboux:2017}, we model this through the use of metal matrices which transform model metal correlations from their correct separations $(\Tilde{r}_{||},\Tilde{r}_\bot)$ to the coordinate grid of the measured correlations:
\begin{equation}
    \xi_\mathrm{m}^M = \sum_N M_{MN} \xi_\mathrm{m}(\Tilde{r}_{||}(N),\Tilde{r}_\bot(N)),
\end{equation}
with the metal matrix:
\begin{equation}
    M_{MN} = \frac{1}{W_M} \sum_{(i,j)\in M, (i,j)\in N} w_i w_j,
    \label{eq:metal_mat}
\end{equation}
where $(i,j)\in M$ refers to bins computed using the assumed (wrong) redshifts, $(i,j)\in N$ refers to bins computed using the correct redshifts, and we compute correlations $\xi_m$ for metal lines $m$ described below. One major change for the DESI DR1 analysis is that we now compute the sum in \Cref{eq:metal_mat} only as a function of $r_{||}$, and ignore the few per cent changes in $r_\bot$. Previous analyses computed these matrices numerically using a small fraction of the pairs available ($\sim0.1\%$), which was not precise enough and very expensive computationally. This simplification allows us to obtain a more precise measurement of the metal contamination when working with the smaller $2$\hMpc\ bins.

We model the contamination from the four metal lines in \Cref{tab:metal_strengths}: SiIII$(1207)$, SiII$(1190)$, SiII$(1193)$, and SiII$(1260)$. The correlation model including metal contamination is given by:
\begin{equation}
    \xi^t_{\mathrm{Ly}\alpha \times X} = \xi_{\mathrm{Ly}\alpha \times X} + \sum_m \xi_{X \times m} + \sum_{m_1,m_2} \xi_{m_1 \times m_2},
\end{equation}
where the sums are performed over the four metal lines, and $X$ stands for \lya\ in the auto-correlation model, and for QSO in the cross-correlation model.

The only missing ingredient in $\xi^t$ is the effect of the distortion due to continuum fitting errors. As discussed in \Cref{subsec:deltas}, we use a projection formalism to account for this effect. This includes forward modelling the projection matrices computed for each forest (\Cref{eq:projection}) into distortion matrices given by:
\begin{align}
    D_{MN}^\mathrm{auto} &= \frac{1}{W_M} \sum_{i,j \in M} w_i w_j \sum_{i',j' \in N} \eta_{ii'} \eta_{jj'}, \\
    D_{MN}^\mathrm{cross} &= \frac{1}{W_M} \sum_{i,j \in M} w_i w_j \sum_{i',j \in N} \eta_{ii'},
\end{align}
with the first equation giving the distortion matrix for the auto-correlation, and the second giving the distortion matrix for the cross-correlation. The model bins $N$ are $2$\hMpc\ in size, while the data bins $M$ are $4$\hMpc\ in size. This means the distortion matrices are not square. The distorted correlation function model is given by:
\begin{equation}
    \hat{\xi}_M = \sum_N D_{MN} \xi_N^t.
\end{equation}

As mentioned above, the final step of our modelling process involves combining the peak and smooth components, which have so far gone through the modelling process independently. The final model is given by:
\begin{equation}
    \xi(r_{||},r_\bot) = \hat\xi_\mathrm{s}(r_{||}, r_\bot) + \hat\xi_\mathrm{p}(\alpha_{||} r_{||}, \alpha_\bot r_\bot),
    \label{eq:xi_sum}
\end{equation}
where $\alpha_{||}$ and $\alpha_\bot$ are scale parameters that we fit for. These correspond to:
\begin{equation}\label{eq:BAO}
    \alpha_{||} = \frac{D_H(z)/r_d}{[D_H(z)/r_d]_{fid}}, \;\;
    \alpha_\bot = \frac{D_M(z)/r_d}{[D_M(z)/r_d]_{fid}},
\end{equation}
where $r_d$ is the scale of the sound horizon at the end of the drag epoch, the $fid$ subscript indicates the values computed in the fiducial cosmology, and $D_H(z) = c/H(z)$ with Hubble parameter $H(z)$ and speed of light $c$. As discussed in \Cref{subsec:mock_boxes}, we use the Planck 2015 results as the fiducial cosmology for all our mocks, which matches the cosmology used to create the \lyacolore\ mocks but is slightly different than the one used for \saclay\ mocks (Planck 2018). We account for this small difference by simply rescaling the \apar\ and \atrans\ results from \saclay\ mocks by a factor given by the ratios between $[D_H(z)/r_d]_{fid}$ and $[D_M(z)/r_d]_{fid}$ in the two cosmologies. This consists of a $\sim0.02\%$ shift in the BAO position. We also test our sensitivity to the choice of fiducial cosmology in \Cref{app:fid_cosmo}.

To obtain posterior distributions for the parameters, we use a Gaussian likelihood and either the minimizer \iminuit\ \cite{iminuit,James:1975dr} or the nested sampler \texttt{PolyChord} \cite{Handley:2015a,Handley:2015b}. \iminuit\ is useful for quickly computing the best-fit model and associated parameter values. However, it approximates parameter uncertainties as Gaussian using the second derivative of the likelihood around the best-fit point. As BAO posteriors can be non-Gaussian \cite{Cuceu:2020}, the more robust method involves computing the full posterior distribution with \texttt{PolyChord}. However, this is much slower and it is not computationally feasible to run the sampler for all 150 mock analyses. Therefore, we only use the sampler to validate the BAO measurements from the stacked correlations, and on one of the mocks to confirm that the Gaussian approximation works well. We found that for DESI DR1 mock datasets, the BAO posterior distribution is closely Gaussian, and therefore the results from the minimizer can be trusted when studying the population of mocks. We use wide flat priors for all parameters, with the exception of the HCD RSD parameter, $\beta_\mathrm{HCD}$, for which we use a Gaussian prior $\mathcal{N}(0.5,0.09^2)$, in line with previous analyses. The priors are the same as the ones used in \kplya.

We show the best-fit model for the stacked \lyacolore\ correlation functions in \Cref{fig:colore_stack}, and for the \saclay\ mocks in \Cref{fig:saclay_stack}. The figures also include the uncertainties from the DESI DR1 data (shaded regions). The stacked correlations are visually well fit by our model, with the model lines generally going through the data points in most regions. For the few exceptions where the model deviates significantly from the data points (e.g. around the metal peak at $60$\hMpc\ in the line-of-sight wedge of auto-correlations), it is still within the shaded region, which indicates that it works well relative to the uncertainties of DESI DR1.

%%%%%%%%%%%%%%%%%%%%%%%%%%%%%%%%%%%%%%%%%%%%%%%%%%%%%%%%%%%%%%%%%%%%%%%%%%%%%%%%%%%%%%%%%%%%%%%%5

\section{Results}
\label{sec:results}

We focus on two types of analyses when fitting correlation functions from synthetic data sets. First, we work with stacks of correlation functions in \Cref{subsec:res_stack}. These allow us to obtain high statistics correlation function measurements, and validate BAO analyses with very high precision. Secondly, we fit each mock individually and study the statistics of the population of resulting BAO measurements in \Cref{subsec:res_pop}. To further test the robustness of our analysis, we study how sensitive BAO measurements are to different covariance matrix estimates in \Cref{subsec:res_cov}, and the impact of redshift errors in \Cref{subsec:res_zerr}.

\subsection{Fits of stacked correlations}
\label{subsec:res_stack}

\begin{figure}
\centering
\includegraphics[width=0.7\linewidth]{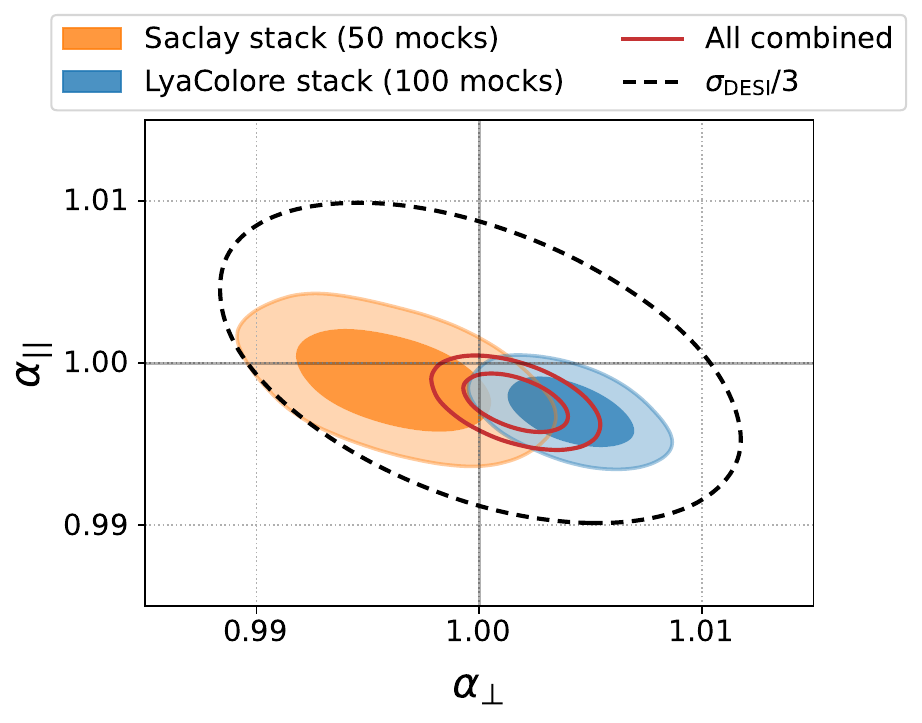}
\caption{$68\%$ and $95\%$ credible regions of BAO measurements from stacks of 50 \saclay\ (orange), 100 \lyacolore\ mocks (blue), and the combination of the two results (red). We measure anisotropic BAO parameterized through the scale parameters \apar\ (along the line-of-sight) and \atrans\ (across the line-of-sight). The grey cross at $\alpha_{||}=\alpha_\bot=1$ indicates the input mock cosmology, while the black dotted contour shows $1/3$ of the DESI DR1 \lya\ BAO uncertainty. The $1/3$ bound represents the threshold within which the analysis had to be validated.  The mock results are consistent with the truth which indicates any bias in the BAO peak position is negligible relative to our uncertainties.}
\label{fig:stack}
\end{figure}

The main goal of this work is to validate the DESI DR1 \lyaf\ BAO measurement using synthetic datasets. To achieve this, we first focus on extracting BAO from stacked correlation functions. We work with the two types of mocks (\lyacolore\ and \saclay) independently. This means we have two sets of measurements for the four correlation functions: the first from 100 \lyacolore\ mocks, and the second from 50 \saclay\ mocks. As we used the same cosmology both to create and to analyse the mocks, we expect to recover $\alpha_{||}=\alpha_\bot=1$ in the absence of systematic errors.

The analysis presented here was performed in parallel with the analysis of DESI DR1 data presented in \kplya. One of the key requirements for unblinding the data measurement was to recover unbiased BAO measurements from stacks of many mocks that include all of the main \lyaf\ contaminants. Concretely, this meant the best-fit measurements from the stacks of mocks needed to be close to the truth to within a certain threshold. This threshold was based on the uncertainty of the DESI DR1 \lya\ BAO measurement, and it required the total systematic bias to be smaller than $1/3$ of that uncertainty.\footnote{The $1/3$ factor is motivated by the fact that adding a systematic error of that magnitude in quadrature would lead to a $\sim5\%$ change in uncertainty.} For the blinded data this corresponded to $\sim0.005$ in \apar\ and $\sim0.007$ in \atrans. After unblinding the uncertainties increased slightly, and the new threshold corresponds to $\sim0.007$ in \apar\ and $\sim0.008$ in \atrans\ (\kplya). For this manuscript, we show the threshold based on the real unblinded measurement, but note that the tighter requirement based on the blinded data had to be satisfied for the unblinding to take place.

\begin{table}[]
    \centering
    \begin{tabular}{c|c|c|c}
        % \hline
         & \lyacolore\ stack & \saclay\ stack & Combined \\
        % \hline
        Parameter & (100 mocks) & (50 mocks) & (150 mocks) \\
        \hline%\hline
        $\alpha_{||}$ & $0.9970\pm 0.0014$ & $0.9992\pm 0.0021$ & $0.9976\pm 0.0012$ \\
        % \hline
        $\alpha_\bot$ & $1.0041\pm 0.0018$ & $0.9964\pm 0.0029$ & $1.0016\pm 0.0015$ \\
        % \hline
        $\rho_{\alpha_{||},\alpha_\bot}$ & -0.49 & -0.47 & -0.48 \\
        % \hline
    \end{tabular}
    \caption{BAO best fit results (mean of posterior), uncertainties ($68\%$ credible regions), and correlation coefficient $\rho$, measured from stacked correlation functions.}
    \label{tab:stack_res}
\end{table}

We show the BAO measurements from the two stacks of mocks in \Cref{fig:stack} (blue and orange contours), and \Cref{tab:stack_res}. Both measurements contain the truth within their $2\sigma$ bound. This indicates there are no systematic effects that significantly bias BAO measurements from our mocks. The red contour shows the combination of the two measurements at the BAO level,\footnote{This combination is done using the Gaussian posteriors.} which is also unbiased. The threshold mentioned above is shown through the black dashed contour in \Cref{fig:stack}, which marks $1/3$ of the $68\%$ credible region of the data BAO measurement from \kplya.\footnote{Computed by rescaling the Gaussian covariance of \apar\ and \atrans.} Both the best-fit results and the entire $95\%$ credible regions of our measurements are within this bound.

%%%%%%%%%%%%%%%%%%%%%%%%%%%%%%%%%%%%%%%%%%%%%%%%%%%%%%%%%%%%%%%%%%%%%%%%%%%%%%%%%%%%%%%%%%%%%%%%5

\subsection{Population statistics}
\label{subsec:res_pop}

\begin{figure}
\centering
\begin{subfigure}{.5\textwidth}
    \centering
    \includegraphics[width=1.0\textwidth,keepaspectratio]{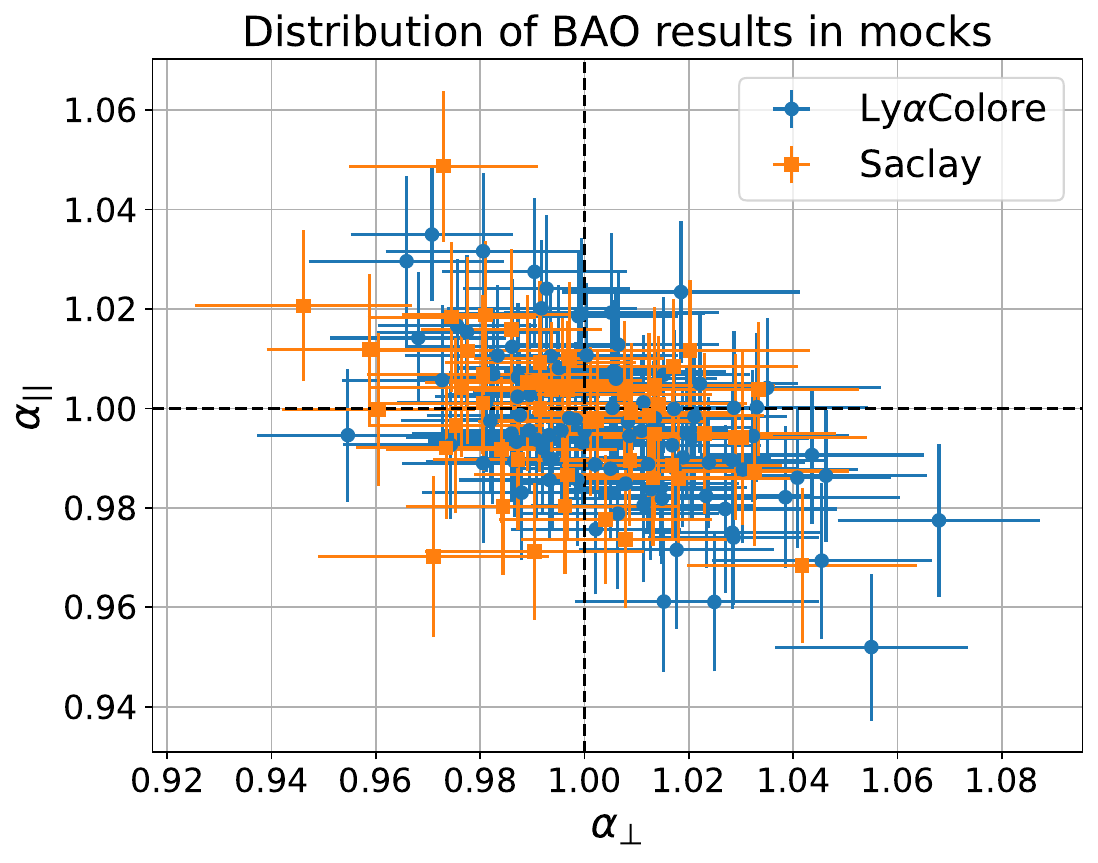}
\end{subfigure}%
\begin{subfigure}{.5\textwidth}
    \centering
    \includegraphics[width=1.0\textwidth,keepaspectratio]{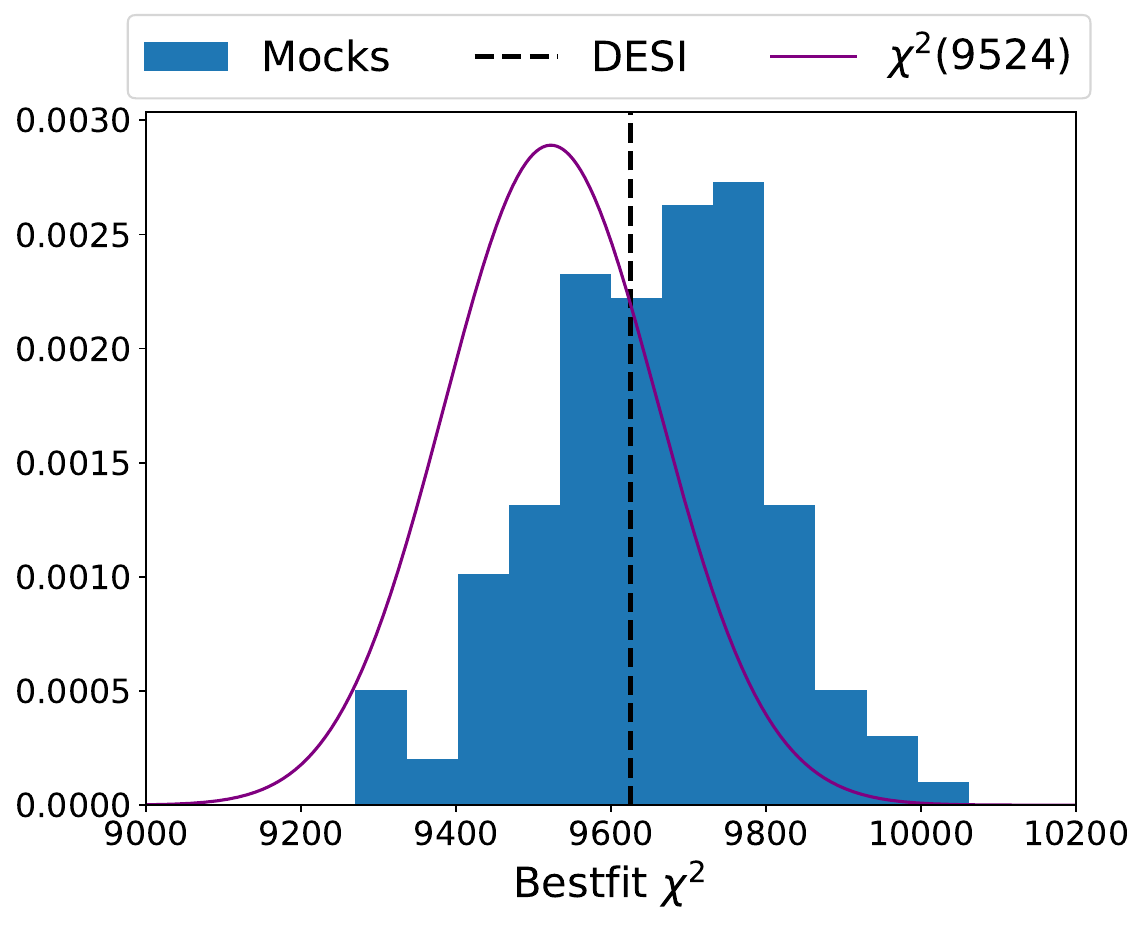}
\end{subfigure}
\caption{(left) Individual BAO measurements from 100 \lyacolore\ (blue) and 50 \saclay\ (orange) mocks. These results are obtained using the \iminuit\ minimizer. The black dashed lines at $\alpha_{||}=\alpha_\bot=1$ indicate the input mock cosmology. (right) Histogram of the best-fit $\chi^2$ statistic (blue), along with the expected $\chi^2$ distribution (purple line), and the best-fit $\chi^2$ value obtained from the real DESI data (black dashed line). The data value is consistent with both the distribution of mocks and the expected distribution. However, the distribution of mocks is systematically shifted to larger values compared to the expected $\chi^2$ distribution. We discuss this shift in \Cref{app:monte_carlo}.}
\label{fig:mock_pop}
\end{figure}

We next shift our focus to analysing the mocks individually and studying the statistics of the entire mock population. The individual BAO measurements from the 100 \lyacolore\ and 50 \saclay\ mocks are shown in the left panel of \Cref{fig:mock_pop}. They are scattered around the truth ($\alpha_{||}=\alpha_\bot=1$), with the two distributions of \lyacolore\ and \saclay\ mocks showing similar variance. We have checked that the two populations are consistent with each other by first performing all the tests described below separately for \lyacolore\ and \saclay\ mocks. As we found no indication of inconsistency, we combine the two sets of mocks and present the statistics of all 148 mock BAO measurements below.

The distribution of best-fit $\chi^2$ values is shown in the right panel of \Cref{fig:mock_pop}, along with the best-fit $\chi^2$ value obtained from the DESI DR1 data. The expectation for mocks is to recover a $\chi^2$ distribution with $9540-16=9524$ degrees of freedom. For data, \kplya fit 17 parameters resulting in a slightly smaller $9523$ degrees of freedom. The $\chi^2$ value obtained from DESI DR1 (black dashed line) is consistent with both the distribution of mock values and the expected $\chi^2$ distribution. However, we find that the distribution of best-fit $\chi^2$ values in our mocks is somewhat larger than the expected value, having a mean and standard deviation of $9654\pm148$. This shift to larger $\chi^2$ values is caused for the most part by the failure of our linear theory model to fit the correlation functions of the mocks across the entire range of scales we use.\footnote{The noisy estimates of the covariance matrix also play a role in this shift as discussed in \Cref{subsec:res_cov}, but they can only explain at most $20\%$ of the shift.} We demonstrate this using Monte Carlo simulations of correlation functions in \Cref{app:monte_carlo}. This result is not surprising given that our mocks deviate from linear theory at small scales (quite significantly in the case of \lyacolore\ mocks). The analysis of \cite{Farr:2020} had to use a significantly narrower range of $40<r<160$ \hMpc\ to obtain a good fit to correlation functions measured directly from the \lyacolore\ transmitted flux boxes. We aim to stay as close as possible to the analysis on real DESI data, which means we use the extended $10<r<180$ \hMpc\ range. Therefore, we expect the linear theory model to not be able to accurately fit this entire range. Furthermore, the failure of the linear theory model extends to large scales due to the metal peaks present along the line-of-sight in our correlation functions, as these peaks represent the small-scale cross-correlation between \lya\ flux (or quasars) and the individual metal absorbers (see \Cref{fig:colore_stack,fig:saclay_stack}). Despite these failures of our model, we are still able to obtain unbiased measurements of the BAO peak position as shown above. Furthermore, while our goal here is to study BAO fits using a physical model, \kplya also tested modelling variations with added broadband polynomials which effectively marginalize over potential systematic effects caused by the failure of the model on small scales.

\begin{figure*}
\includegraphics[width=1.0\linewidth]{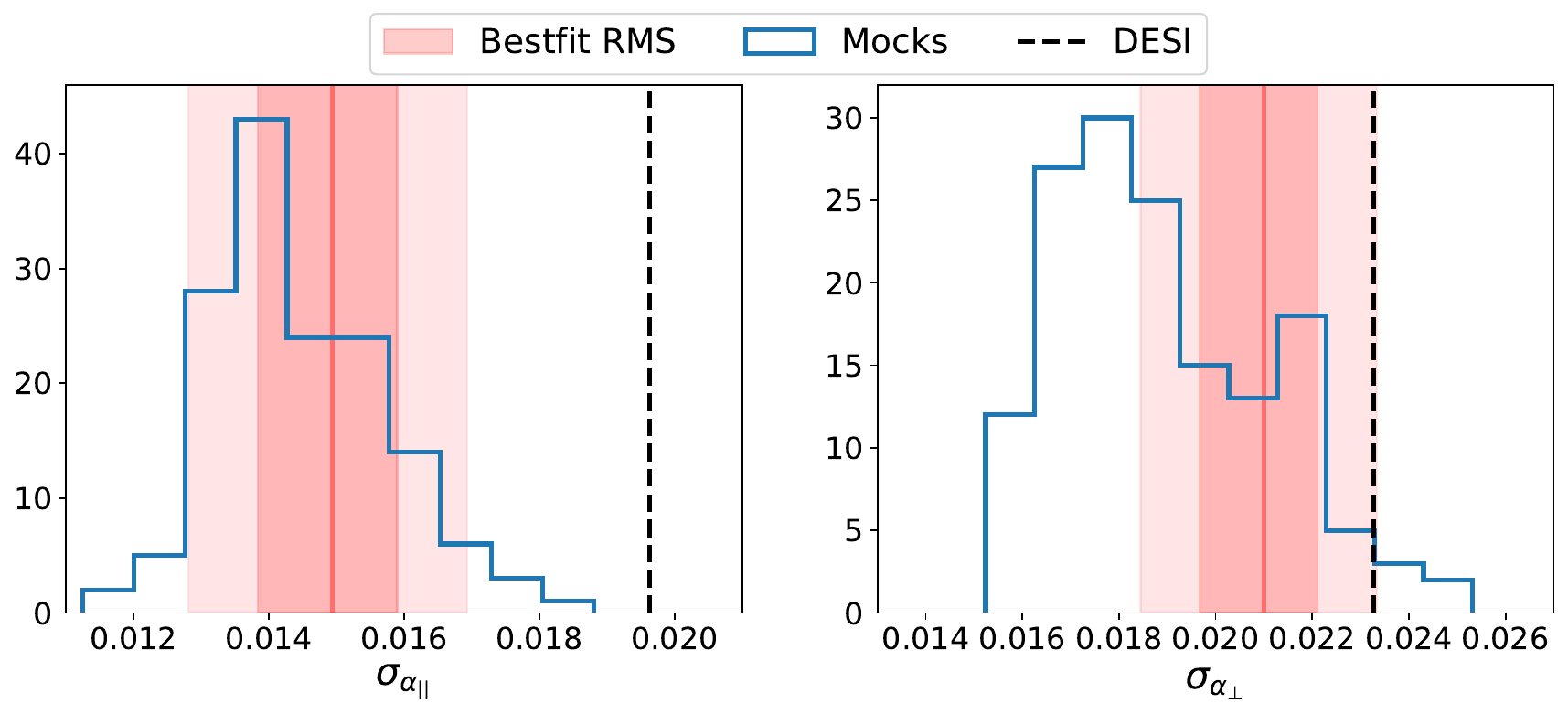}
\caption{Distribution of uncertainties for the 150 BAO measurements from mocks. Uncertainties in \apar\ are shown in the left panel, and uncertainties in \atrans\ are shown in the right panel. The red line and bars indicate the scatter of BAO best-fit measurements in the population of mocks, with the darker (lighter) region indicating the $68\%$ ($95\%$) credible region. This is consistent with the blue histogram, indicating the measured uncertainties are representative of the real uncertainty in these measurements. The dashed black lines indicate the uncertainties in the DESI DR1 \lya\ BAO measurement.}
\label{fig:sigma_distrib}
\end{figure*}

Another important test that synthetic datasets allow us to perform is checking the recovered BAO uncertainties, $\sigma_{\alpha_{||}}$ and $\sigma_{\alpha_\bot}$, against the scatter of BAO best fits. Assuming Gaussian distributed results, if the covariance matrix estimation is unbiased and we correctly marginalize all nuisance parameters, we expect the two to be consistent with each other. We show the distribution of uncertainties on \apar\ and \atrans\ in \Cref{fig:sigma_distrib}, along with the RMS deviation of best-fit values (red line and bars). We find that the scatter of BAO best fits in mocks is broadly consistent with the reported uncertainties (red line is within the blue histogram). As we have a fairly limited number of mocks (150), this comparison is quite noisy as shown by the large uncertainties on the RMS deviation of best-fit values. We test our uncertainties more rigorously using the pull distributions below. In \Cref{fig:sigma_distrib}, we also show the uncertainties obtained from DESI DR1 (black line). This appears to be quite an extreme result when compared to the mocks, with only $5$ mocks having a larger \atrans\ uncertainty, and no mocks having an uncertainty as large as the data in \apar. However, this difference can be explained by the fact that our mocks do not include the effect of BAO broadening due to non-linear evolution. We demonstrate this in \Cref{sec:discussion} using Monte Carlo simulations.

\begin{figure*}
\includegraphics[width=1.0\linewidth]{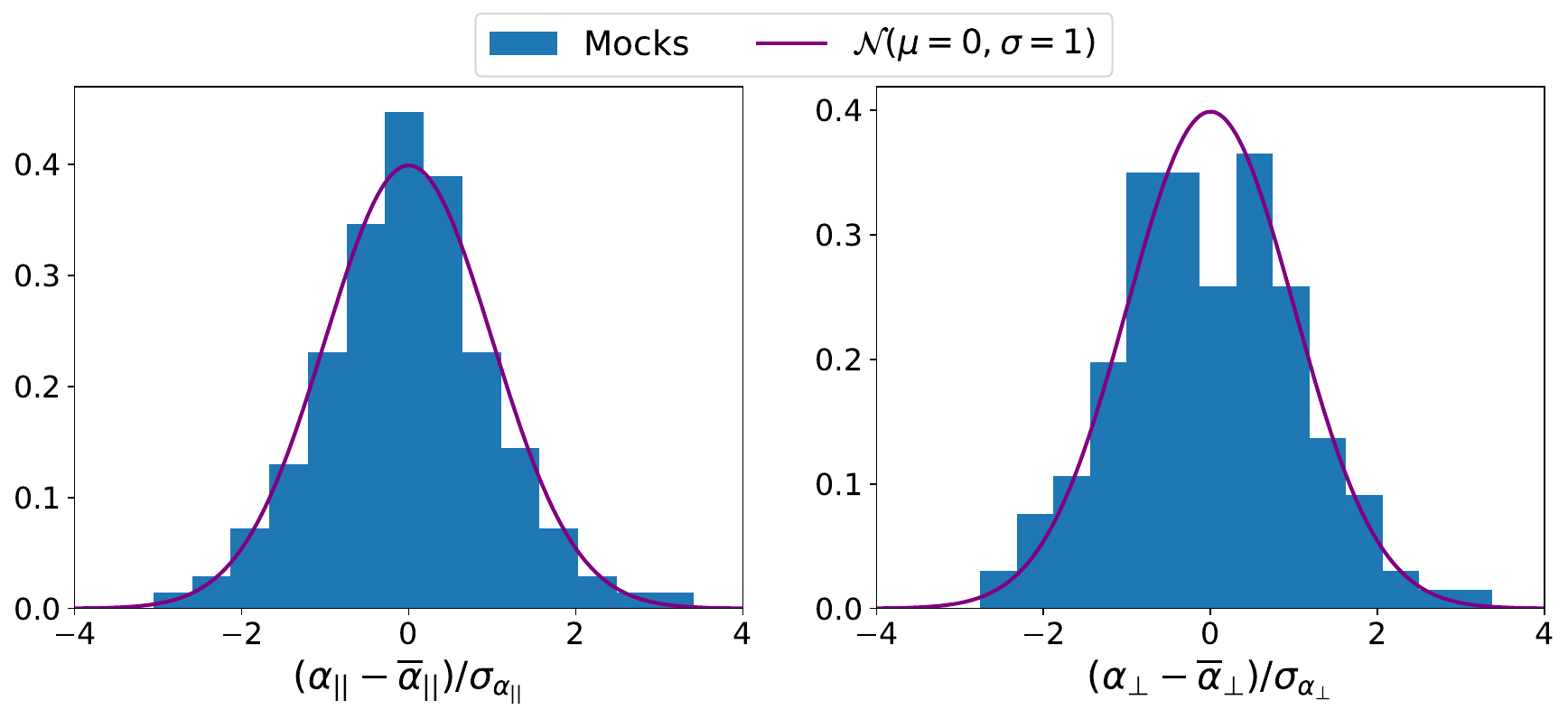}
\caption{Distribution of normalized residuals in \apar\ (left) and \atrans\ (right) from the set of 150 BAO measurements in mocks. The purple lines indicate the expected distributions, i.e. Gaussian with unit variance.}
\label{fig:pull_distrib}
\end{figure*}

As a final test of the measured uncertainties, we show the pull distributions for \apar\ and \atrans\ in \Cref{fig:pull_distrib}. These are obtained by subtracting the mean value of mock BAO best fits from each individual result, and dividing by the measured uncertainty (i.e. $[\alpha-\Bar{\alpha}]/\sigma_{\alpha}$). We can test whether our measured uncertainties are representative of the real uncertainty by checking if the standard deviations of the pull distributions are consistent with unity. \Cref{fig:pull_distrib} includes a unit variance Gaussian for comparison. The standard deviations of these distributions are $1.02\pm0.07$ for $\Delta\alpha_{||}/\sigma_{\alpha_{||}}$, and $1.11\pm0.06$ for $\Delta\alpha_\bot/\sigma_{\alpha_\bot}$, with uncertainties obtained through bootstrap. This test assumes both that the distribution of BAO best-fits is Gaussian (through the use of the standard deviation), and that the posterior distributions of individual measurements can be approximated as Gaussian (because we use Gaussian uncertainties from \iminuit). Neither of these were true for previous \lyaf\ BAO analyses \cite[e.g.][]{duMasdesBourboux:2017,Bautista:2017,duMasdesBourboux:2020}. However, in the case of DESI DR1 mocks, \Cref{fig:pull_distrib} shows that the Gaussian assumption works very well for \apar, and moderately so for \atrans. If instead of the standard deviation, we use the $68\%$ credible region,\footnote{This is done by computing half the distance between the 16th and 84th percentiles.} we obtain $0.93\pm0.09$ for $\Delta\alpha_{||}/\sigma_{\alpha_{||}}$, and $1.08\pm0.09$ for $\Delta\alpha_\bot/\sigma_{\alpha_\bot}$. This does not rely on the first assumption of Gaussian distributed best-fits, and therefore confirms both that the second assumption is justified (we can approximate posterior distributions as Gaussian), and that the uncertainties are well estimated using our framework. We end by noting that while this analysis validates our BAO uncertainties, it only does so at the $\sim9\%$ level, which is expected given the relatively small number of mocks we have (150).

%%%%%%%%%%%%%%%%%%%%%%%%%%%%%%%%%%%%%%%%%%%%%%%%%%%%%%%%%%%%%%%%%%%%%%%%%%%%%%%%%%%%%%%%%%%%%%%%5

\subsection{Covariance matrix tests}
\label{subsec:res_cov}

\begin{figure*}
\includegraphics[width=1.0\linewidth]{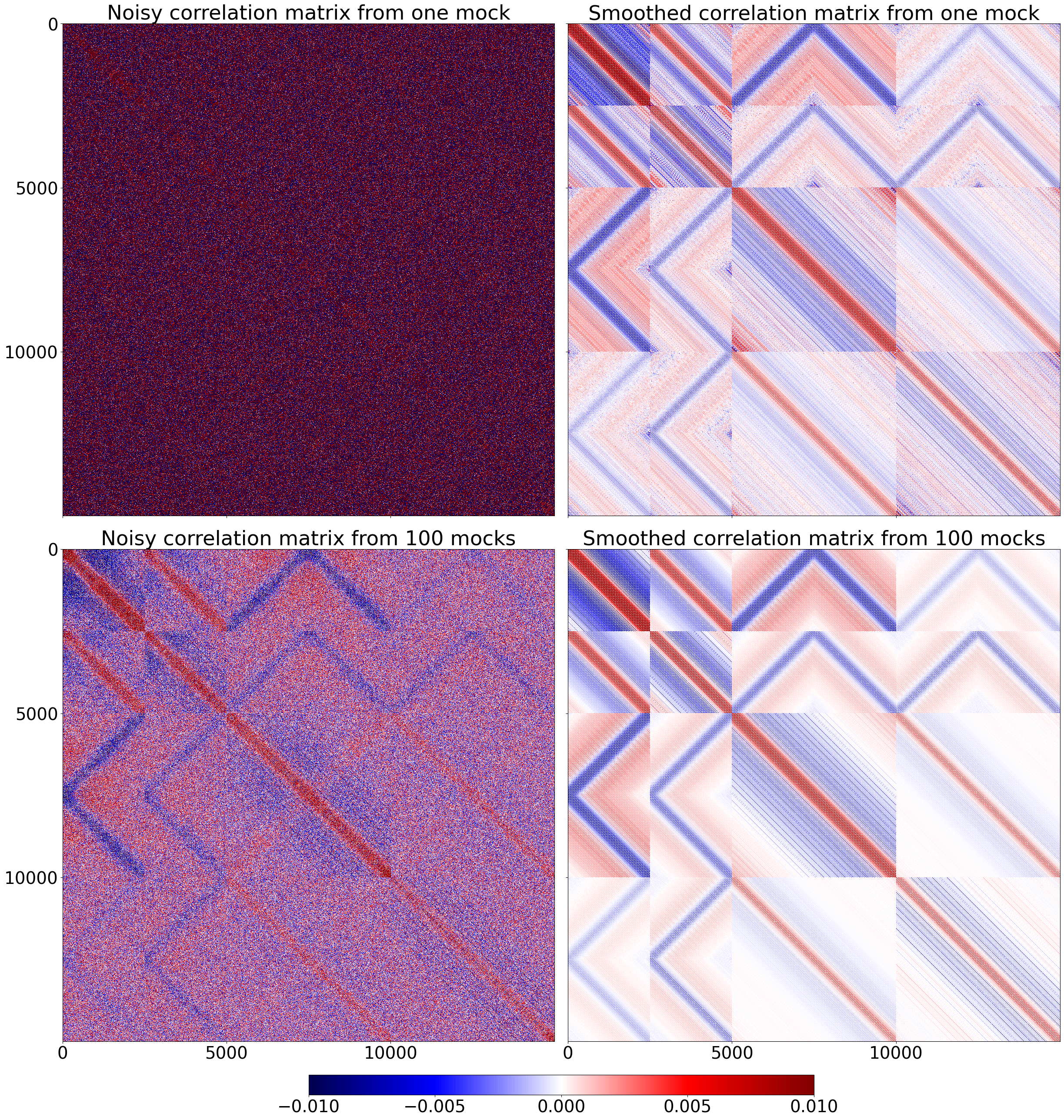}
\caption{Global correlation matrices of all four \lyaf\ correlation functions. The top row shows correlation matrices computed from one \lyacolore\ mock, while the bottom row shows the mean correlation matrices computed from all 100 \lyacolore\ mocks. We plot the initial noisy estimates of the correlation matrix in the left column, while the smoothed versions are shown in the right column. In each panel, the global correlation matrix is made up of the individual auto-correlation matrices for each of the four correlation functions in order: \lyalyalyalya, \lyalyalyalyb, \lyalyaqso, \lyalybqso\ (diagonal blocks), and all the possible cross-correlation matrices (off-diagonal blocks). Note that the range of the colour scale is 100 times smaller than in \Cref{fig:correlation_matrix_auto}.}
\label{fig:correlation_matrices}
\end{figure*}

The DESI DR1 \lya\ BAO analysis uses one global covariance matrix for all four \lya\ correlation functions. This is because, for this data set, the cross-covariance between the different correlation functions cannot be ignored as was done with previous BOSS and eBOSS analyses. This is discussed in detail in \kplya. The result of needing to account for the cross-covariance is that we now need to estimate a $15000 \times 15000$ covariance matrix. Due to computational and storage constraints, we are limited to a sample of 150 mock data sets, which means we cannot use mocks to directly estimate our covariance matrix. Therefore, for this work, we rely on the same method used in previous \lya\ BAO analyses to estimate our covariance matrices \cite{Delubac:2015,duMasdesBourboux:2020}. As described in \Cref{subsec:cov_mat}, this involves first computing a noisy estimate from the $1028$ correlation function measurements in $250\times250 \;(h^{-1}\mathrm{Mpc})^2$ patches on the sky, and then smoothing this noisy estimate to obtain our measurement of the covariance matrix. This process was used to estimate the data covariance matrix in \kplya, and the covariance matrices for each of our mocks here. The method has also been tested against the much slower method of computing the Gaussian covariance using the Wick approximation, and the two were found to produce consistent results \cite{duMasdesBourboux:2020}. In this section, we wish to use the population of mocks to study how well this method works for individual mocks, and to test the impact on BAO measurements.

In \Cref{subsec:cov_mat} we also introduced the covariance matrix estimates from the stack of mocks (which were used to obtain the results in \Cref{subsec:res_stack}). These estimates use $\sim100$k samples for \lyacolore\ and $\sim50$k samples for \saclay\ instead of the $1028$ we have in individual mocks. Therefore, they represent much better estimates of the covariance matrices of these mock data sets. We show the normalized global covariance matrices (i.e. correlation matrices) from one \lyacolore\ mock and from the stack of 100 \lyacolore\ mocks in \Cref{fig:correlation_matrices}. Each of these global matrices contains the individual auto-correlation matrices (diagonal blocks) for the four correlation functions in order: \lyalyalyalya, \lyalyalyalyb, \lyalyaqso, \lyalybqso. The two auto-correlation functions have covariance matrices of size $2500\times2500$, while the two cross-correlation functions have covariance matrices of size $5000\times5000$. Off-diagonal blocks represent the cross-covariance matrices between different correlation functions. The strongest of these is the cross-covariance between \lyalyalyalya\ and \lyalyaqso\ (\Cref{fig:correlation_matrix_auto_cross}).

\begin{figure}
\centering
\includegraphics[width=1\linewidth]{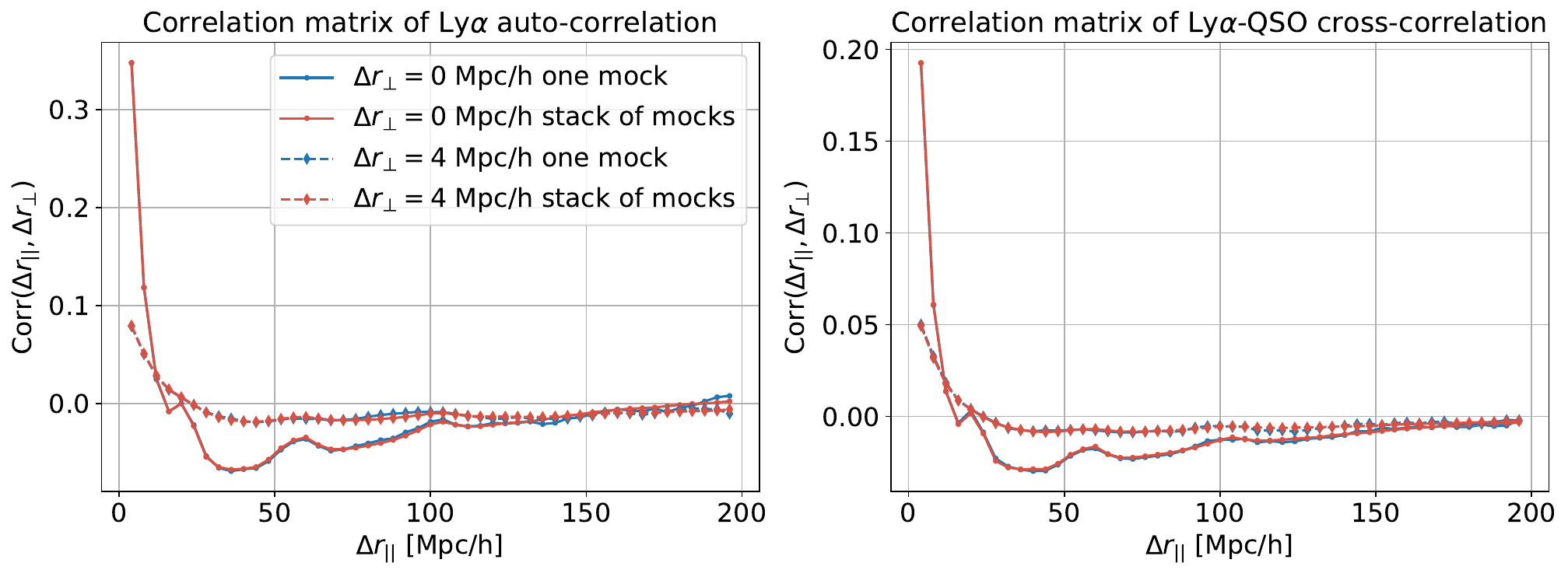}
\caption{The smoothed normalized covariance matrix (correlation matrix) for the \lya\ auto-correlation (left) and the \lya-QSO cross-correlation (right). We compare the smoothed correlation matrix as a function of line-of-sight separation for one \lyacolore\ mock versus the stack of 100 \lyacolore\ mocks. This figure shows that the smoothing applied to the very noisy estimate of the covariance from one mock leads to a correlation structure that is remarkably similar to the smoothed correlation matrix from the stack of mocks.}
\label{fig:corr_mat_smoothing}
\end{figure}

From \Cref{fig:correlation_matrices} we can see that the initial estimate of the covariance matrix in individual mocks (top left) shows no clear features away from the diagonal due to noise. We can start to see the features of the correlation matrix when looking at the initial estimate from the stack of mocks (bottom left) due to the much larger sample size used to compute it. However, this estimate is still fairly noisy, which is why we also apply the smoothing procedure in this case. The two panels on the right show the smoothed correlation matrices from one mock (top) and from the stack of mocks (bottom). These two show that the smoothed covariance matrix from the stack of correlation functions is remarkably similar to the smoothed covariance estimated from individual mocks. This consistency is better illustrated in \Cref{fig:corr_mat_smoothing}, which shows the line-of-sight components of the covariance matrix for the \lya\ auto-correlations and the \lya-QSO cross-correlation. We focus on correlations between line-of-sight bins as they are the most important part of the covariance matrix (see \Cref{subsec:cov_mat}).

As the initial estimates of the covariance matrix in individual mocks are dominated by noise, we wish to test our method for estimating and smoothing covariance matrices by comparing the results obtained in \Cref{subsec:res_pop} with results obtained with the more robust covariance matrix from the stack of mocks. To do this, we build new covariance matrices from the correlation matrix estimate of the stacks of mocks and the variance estimates in each mock:
\begin{equation}
    C_{MN}^{test} = (C_{MM}C_{NN})^{1/2} Corr_{MN}^\mathrm{stack},
    \label{eq:corr_mat_test}
\end{equation}
where $Corr_{MN}^\mathrm{stack}$ is the correlation matrix estimate from the stack of mocks (given by \Cref{eq:corr_matrix}), and $C_{MN}^{test}$ are our new test covariance matrices for each mock. As two of the mocks were already using these covariance matrices (see \Cref{subsec:cov_mat}), we discard them from the analysis in this section and work with the remaining 148 mocks.

\begin{figure}
\centering
\includegraphics[width=0.6\linewidth]{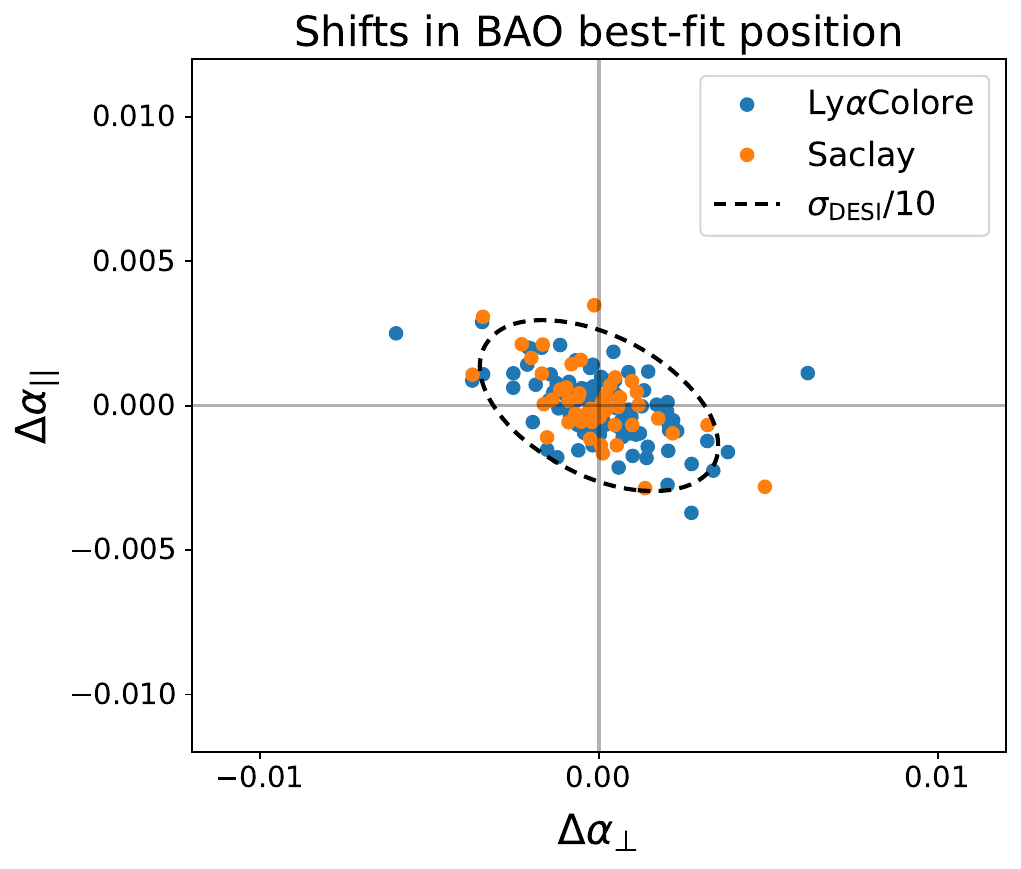}
\caption{Shifts in the BAO best-fit positions from mocks when using the correlation matrix estimated from the stack of all mocks versus the correlation matrix estimated from each individual mock. The stack of mocks provides a much larger sample ($100$k for \lyacolore, and $50$k for \saclay) to estimate the covariance matrix. As shown here, using these improved estimates leads to small random shifts in the measured BAO position. However, these are unbiased and much smaller than the DESI DR1 BAO uncertainty. The black dashed contour indicates one-tenth of the DESI DR1 uncertainty.}
\label{fig:cov_bao_shifts}
\end{figure}

We then fit the mocks using these new estimates and compare the BAO results with the previous results from \Cref{subsec:res_pop}. The difference between the best-fit \apar\ and \atrans\ of each mock is shown in \Cref{fig:cov_bao_shifts}. The shifts in the BAO position are randomly distributed with very small RMS compared to the uncertainty of the DESI DR1 constraints. We measure the RMS of $\Delta \alpha_{||}$ to be $0.0012 \pm 0.0001$, and the RMS of $\Delta \alpha_\bot$ to be $0.0014 \pm 0.0002$.\footnote{Uncertainties are obtained through bootstrap.} These represent less than a tenth of the DESI DR1 uncertainty. We also compared the uncertainties in \apar\ and \atrans\ obtained with the two covariance matrix estimates, and we found the changes are randomly distributed as well, and of the order of $\sim1\%$ of the DESI DR1 uncertainty. This means the two estimates of the covariance matrix produce consistent BAO results.

On the other hand, when comparing the best-fit $\chi^2$ values of the two populations, we found a small systematic shift to smaller $\chi^2$ values when using the correlation matrix estimated from the stack of mocks. The mean and RMS of the shift are roughly $\Delta \chi^2 = -30 \pm 10$, with $147$ out of $148$ mocks having an improved (smaller) $\chi^2$. This shift is too small to explain the results in \Cref{fig:mock_pop}, but indicates that the noisy estimate of the covariance matrix has a small but significant impact on the quality of the model fits.

%%%%%%%%%%%%%%%%%%%%%%%%%%%%%%%%%%%%%%%%%%%%%%%%%%%%%%%%%%%%%%%%%%%%%%%%%%%%%%%%%%%%%%%%%%%%%%%%5

\subsection{Impact of redshift errors}
\label{subsec:res_zerr}

In line with previous analyses \cite[e.g.][]{FontRibera2013,duMasdesBourboux:2020}, we have so far only considered the impact of redshift errors through their smoothing effect on the \lya-QSO cross-correlation. The redshift errors used here were generated using a Gaussian distribution with dispersion $\sigma_z=400\ \text{km/s}$ ( \Cref{subsec:mock_spec}). See \cite{KP6s4-Bault} for a comparison between redshift errors measured in DESI mocks versus DESI data. \cite{Youles:2022} found that redshift errors can also introduce spurious correlations in both the \lya\ auto- and its cross-correlation with quasars. These spurious correlations appear due to the smearing of emission lines in the forest region, leading to errors in the fitted continuum of each forest. These errors do not average out because quasars are not uniformly distributed (i.e. they are clustered). Therefore, the two ingredients that contribute to these spurious correlations are the smearing of emission lines in the forest regions and the quasar auto-correlation function \cite{Youles:2022}. 

In this section, we wish to study the impact of spurious correlations caused by redshift errors on BAO measurements. We did not include this effect in our baseline mocks because we know based on \cite{Youles:2022} that it is more extreme in our mocks than in reality. This has to do with both ingredients that give rise to this effect. Firstly, the quasar auto-correlation in our mocks is significantly larger on small scales \footnote{As much as double at $\sim15$ \hMpc.} compared to realistic simulations (see Figure 6 of \cite{Youles:2022}). This is because we use the log-normal approximation to draw quasar position in both \lyacolore\ and \saclay\ mocks. Secondly, in our mocks, the smearing of emission lines in the forest region is completely random, but in reality, this smearing is likely smaller because we use emission lines (on the red side of the \lya\ line) to measure the redshift. This means that even if the measured quasar redshift has some error relative to the systemic redshift, it is likely better at predicting the position of emission lines in the forest region than the systemic redshift. Therefore, we decided not to include this effect in our baseline mocks. Nevertheless, we study its impact on BAO measurements with the caveat that our results likely overestimate it.

\begin{figure}
\centering
\begin{subfigure}[b]{1.0\textwidth}
    \centering
    \includegraphics[width=\textwidth]{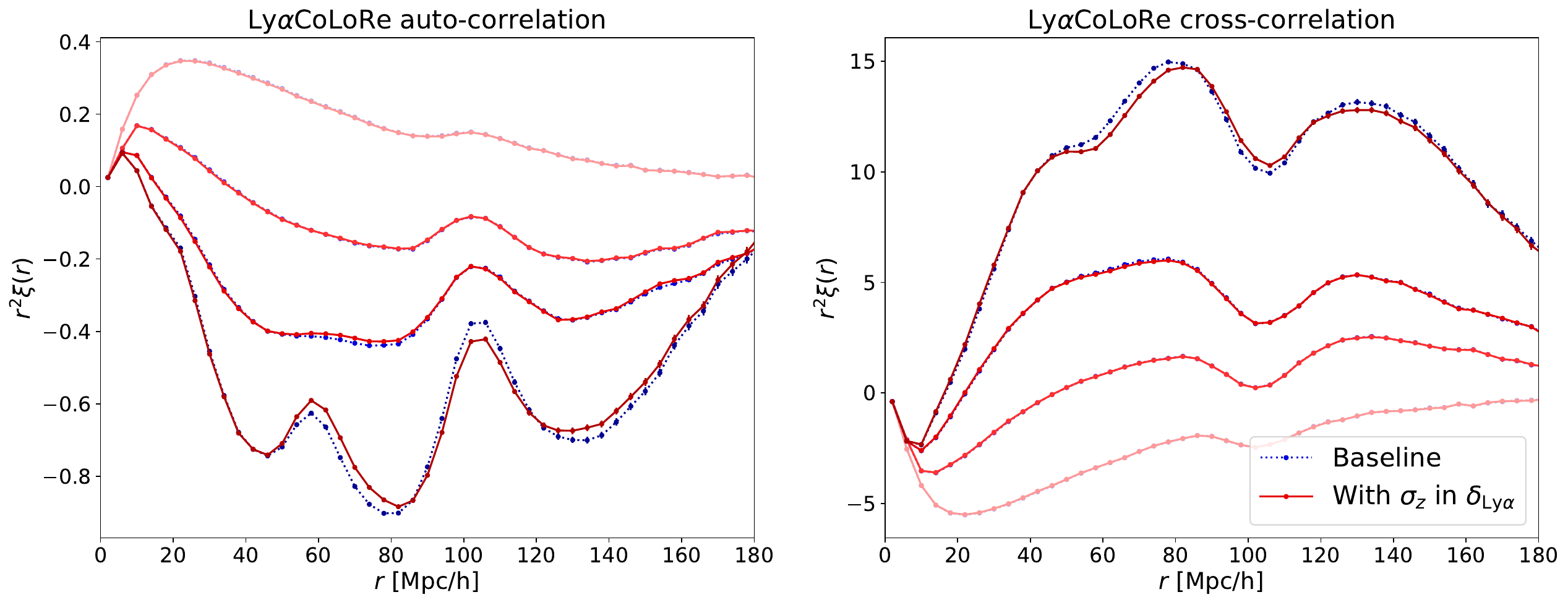}
\end{subfigure}%
\hfill
\begin{subfigure}[b]{1.0\textwidth}
    \centering
    \includegraphics[width=\textwidth]{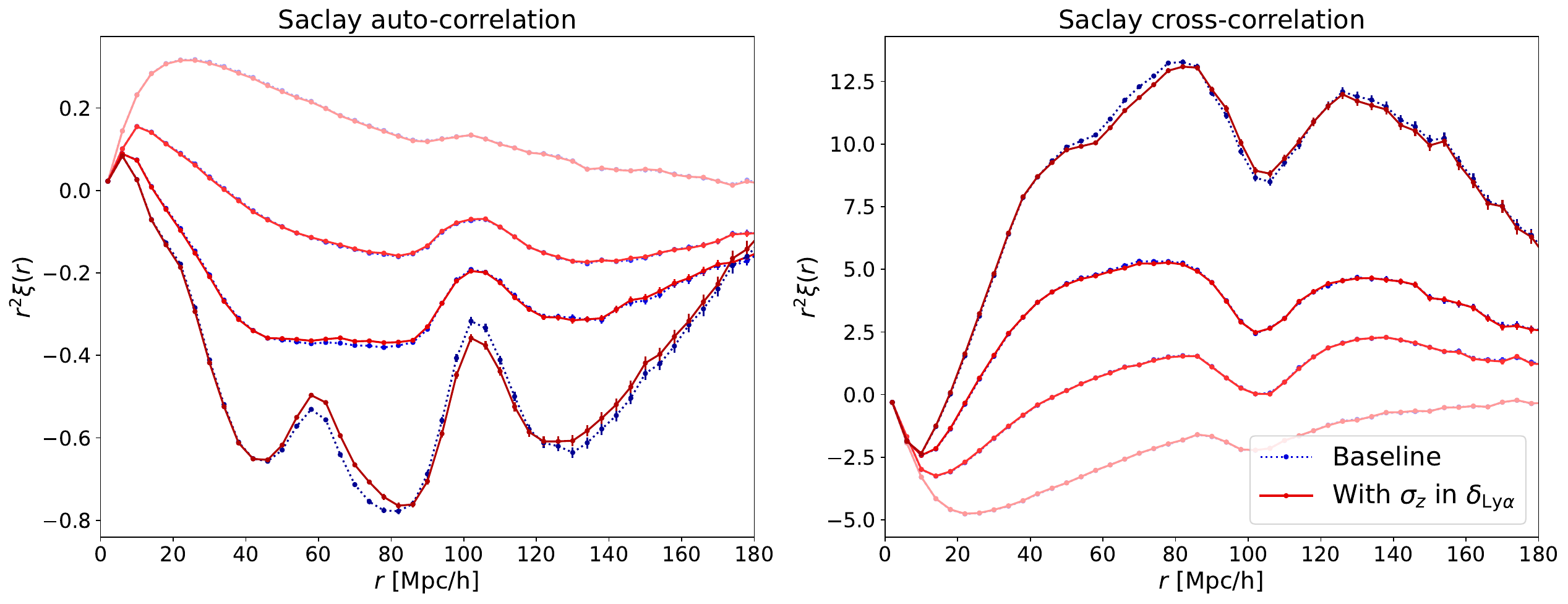}
\end{subfigure}
\caption{Stacked correlation functions compressed into $\mu$ wedges (points with error bars and lines connecting them), with lighter colours indicating smaller $\mu$ values, and darker colours indicating larger $\mu$ values. The top panels show stacked correlations from 100 \lyacolore\ mocks, while the bottom panels show stacked correlations from 50 \saclay\ mocks. The left panels show \lya\ auto-correlations, while the right panels show \lya-QSO cross-correlations. In our baseline mocks (blue), redshift errors only have a smoothing effect on the cross-correlation. On the other hand, red correlations come from mocks where redshift errors also affect the fitted quasar continuum by smearing emission lines in the forest region.
}
\label{fig:zerr_corr}
\end{figure}

We show the impact of the spurious correlations on stacks of correlation functions from \lyacolore\ and \saclay\ mocks in \Cref{fig:zerr_corr}. As shown by \cite{Youles:2022}, these have an impact mostly limited to the line-of-sight wedge, but that extends to very large separations (even larger than the BAO scale). From \Cref{fig:zerr_corr}, this appears most pronounced in the roughly $50-80$ \hMpc\ interval where the blended SiII(1190) and SiII(1193) metal peak is located, and in the roughly $90-110$ \hMpc\ interval that overlaps the location of the BAO peak.

\begin{figure}
\centering
\includegraphics[width=0.7\linewidth]{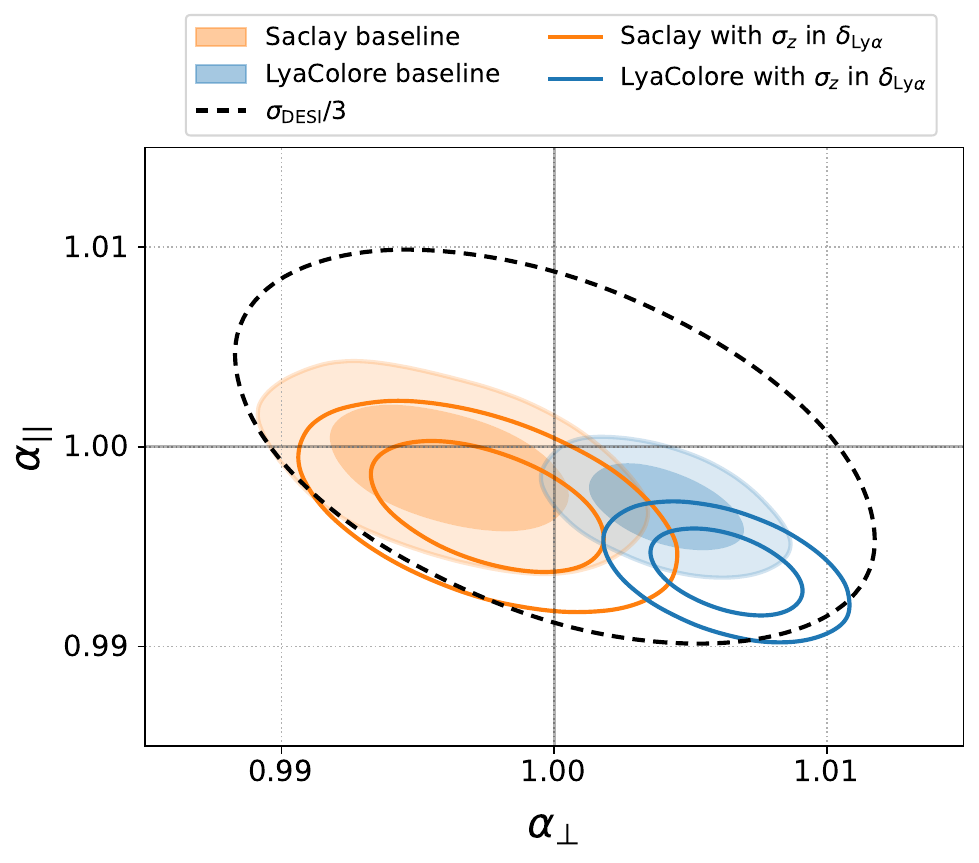}
\caption{Similar to \Cref{fig:stack}, but now including results from mocks where redshift errors affect the fitted quasar continuum. These redshift errors lead to spurious correlations in both the auto- and cross-correlation as shown in \Cref{fig:zerr_corr}. This result indicates that while the spurious correlations have an impact on BAO measurements, this is very small ($<0.2\%$). Importantly, the mock measurements are still within the threshold of $1/3$ of the DESI DR1 BAO uncertainty (blue and orange contours are within the dashed black line).}
\label{fig:res_zerr}
\end{figure}

The presence of these spurious correlations in the BAO region means there could be systematic errors in the measured BAO signal. To check for this, we perform two types of tests. First, we fit the BAO peak position using the stacked correlation functions, similarly to \Cref{subsec:res_stack}, and check if the results are still within the threshold we set above ($1/3$ of the DESI DR1 \lya\ BAO uncertainty). We show these results in \Cref{fig:res_zerr}. While the measurements are slightly different than the results presented in \Cref{fig:stack}, they are both still within the threshold (dashed black line). This indicates the spurious correlations do have an impact on BAO measurements, but the impact is very small.

Secondly, we attempt to quantify the impact on BAO measurements by again fitting all of the mocks (now including spurious correlations due to redshift errors), and computing the mean shift in the BAO position.\footnote{We have also performed this analysis using the median shift instead of the mean, and it did not impact our conclusions.} We find that there is indeed a shift in the BAO positions, which for \lyacolore\ mocks we measure to be $\Delta\alpha_{||}=-0.0032\pm0.0004,\;\Delta\alpha_\bot=0.0027\pm0.0004$, and for \saclay\ mocks we measure as $\Delta\alpha_{||}=-0.0018\pm0.0007,\;\Delta\alpha_\bot=0.0010\pm0.0007$. The quoted uncertainties are derived through bootstrap of the BAO best-fit positions (i.e., these measurements are based on the best-fit BAO positions and not on the posterior distributions).

For this test, it is also useful to re-parameterize our BAO coordinates, into the isotropic BAO component, and the anisotropic component which captures the Alcock-Paczyński (AP) information. Following \cite{Cuceu:2021}, we define the isotropic component as $\alpha=\sqrt{\alpha_\bot \alpha_{||}}$, and the Alcock-Paczyński direction as $\phi=\alpha_\bot/\alpha_{||}$. When working in the $\phi, \alpha$ coordinates, it becomes clear that the systematic shift due to spurious correlations is largely in the Alcock-Paczyński direction, with $\Delta\phi=0.0061\pm0.0008$ for \lyacolore\ mocks and $\Delta\phi=0.0029\pm0.0013$ for \saclay\ mocks. The \lyacolore\ measured bias is very significant ($\sim8\sigma$) and is about double in magnitude compared to the one measured from \saclay\ mocks, which is only $\sim 2.2\sigma$ from $0$. On the other hand, the isotropic BAO measurement does not show the same significant bias, with $\Delta\alpha=-0.0006\pm0.0004$ for \lyacolore\ mocks, and $\Delta\alpha=-0.0008\pm0.0008$ for \saclay\ mocks. The two measurements are now consistent and within $2\sigma$ of $0$. For comparison, the DESI DR1 \lya\ BAO uncertainties, are $\sim0.04$ in $\phi$ and $\sim0.016$ in $\alpha$.

As discussed in \Cref{subsec:mock_boxes}, \saclay\ mocks have more realistic quasar clustering when compared to \lyacolore\ mocks, in which the small-scale quasar auto-correlation is much larger than measured in simulations and real data. This explains why the shift in the AP direction is roughly double in \lyacolore\ mocks compared to \saclay\ mocks. However, the isotropic BAO shifts are consistent with each other and not significant (below $2\sigma$). Therefore, using our sample of mocks, we have strong indications of a bias in the AP direction, but the isotropic BAO direction is not significantly biased at the level tested. Nevertheless, it is clear that these systematic shifts are small compared to the DESI DR1 BAO uncertainties. Taking the \saclay\ results as the more realistic ones, the shifts are less than a tenth of $\sigma_\mathrm{DESI}$ in both directions and as discussed above, our analysis likely overestimates the impact of this effect.

While the spurious correlations studied here have minimal impact on the recovered BAO peak position, they have a much larger impact on other nuisance parameters. The systematic biases are not relevant for BAO, because \kplya found that none of the nuisance parameters are correlated with \apar\ or \atrans.\footnote{Also see \cite{Cuceu:2020} for a discussion on correlations between BAO and nuisance parameters.} However, we mention the most important of these shifts here for completeness. We report all systematic shifts relative to their DESI DR1 uncertainty. We find that both the \lya\ bias ($b_{\mathrm{Ly}\alpha}$) and RSD ($\beta_{\mathrm{Ly}\alpha}$) parameters are shifted to lower values by $\sim1\sigma$ in \lyacolore\ mocks, and $\sim0.5\sigma$ in \saclay\ mocks. The HCD bias ($b_\mathrm{HCD}$) and typical length scale ($L_\mathrm{HCD}$) are both shifted to larger values by $\sim1\sigma$. The most affected parameter is the bias of the SiII$(1190)$ line which is shifted by $\sim2\sigma$ to more negative values. Given the position of this metal peak is at $\sim60$\hMpc, this shift indicates that the metal model is fitting the spurious correlation observed in \Cref{fig:zerr_corr} at the same scales. That is also true for the SiII$(1260)$ peak at $\sim103$\hMpc, but its bias parameter only displays a $\sim0.5\sigma$ shift to larger (less negative) values. As the spurious correlations display similar patterns to the metal contamination (they appear as peaks and troughs along the line-of-sight only, and rapidly decay across the line of sight), it is not surprising that the metal models also fit these features. However, given our results here, creating a model for these spurious correlations so they can be properly marginalized should be one of the top priorities for future \lyaf\ analyses.

%%%%%%%%%%%%%%%%%%%%%%%%%%%%%%%%%%%%%%%%%%%%%%%%%%%%%%%%%%%%%%%%%%%%%%%%%%%%%%%%%%%%%%%%%%%%%%%%5

\section{Discussion}
\label{sec:discussion}

The results presented in this publication were used to validate the DESI DR1 \lyaf\ BAO measurement from \kplya. We performed the same analysis as \kplya on a set of 150 mock data sets that include the major \lyaf\ contaminants and found no significant systematic biases in the recovered BAO peak position. We also found that the BAO uncertainties measured in mocks are consistent with the spread of BAO best-fit measurements and that our method for estimating the covariance matrix leads to consistent results when compared to the improved estimate from a large population of mocks. In this section, we wish to discuss the applicability of our results given the limitations we have in terms of our synthetic \lyaf\ data sets.

The largest limitation with the current generation of \lyaf\ mocks is that they are based on the log-normal approximation. The \lyaf\ mostly probes linear and mildly non-linear regimes due to the self-censoring of large overdensities which quickly absorb all \lya\ flux. Furthermore, DESI measurements of the \lya\ forest are limited to redshifts $z>1.95$, when the matter in the Universe was still linear on much smaller scales than today \cite{Arinyo:2015,Givans2022}. This means that modelling the large scales considered in \lyaf\ BAO analyses only requires small deviations from linear theory \cite{Givans2022,Hadzhiyska2023}. Therefore, log-normal mocks can be very useful, especially when studying the \lya\ auto-correlation, where the main limitation only comes from the fact that these mocks do not include the BAO broadening due to non-linear evolution (see discussion below). When modelling real data, small-scale non-linearities are usually modelled using an ad-hoc equation tuned from hydrodynamical simulations \cite{Arinyo:2015}. However, we found that the presence or absence of this model has no impact on current BAO measurements, both with real data (\kplya), and with the mocks here.

While the log-normal approximation works well for the \lyaf\ on large scales, it fails to correctly simulate the quasar distribution, especially on small scales, which could impact measurements of the \lya-QSO cross-correlation. For the mocks used in this work, this is a larger issue in \lyacolore\ mocks, which greatly overpredict the small-scale quasar clustering \cite{Ramirez-Perez:2022,Youles:2022}. However, this is less of a problem in \saclay\ mocks, because they draw quasars from a simulated quasar density field instead of the matter field, which leads to more accurate small-scale QSO clustering \cite{Etourneau:2023}. This is why we have compared the results from \lyacolore\ and \saclay\ mocks throughout this article. The fact that the main results in \Cref{subsec:res_stack,subsec:res_pop} do not show significant discrepancies between the two types of mocks indicates that this effect does not have a large impact on BAO measurements. The only place where we have found discrepant results was in \Cref{subsec:res_zerr}, where we know the spurious correlations due to redshift errors are directly related to the strength of the quasar clustering \cite{Youles:2022}.

\begin{figure*}
\includegraphics[width=1.0\linewidth]{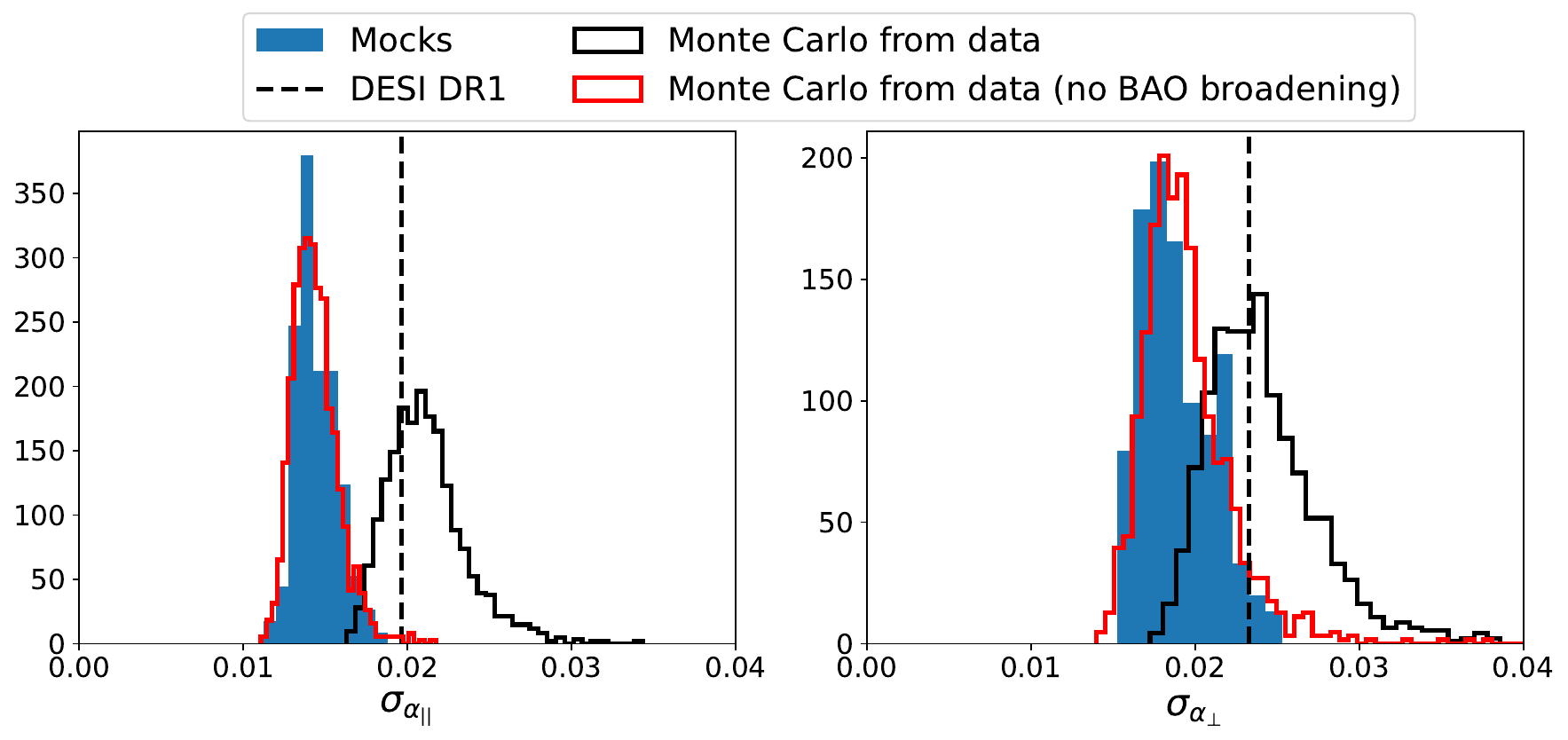}
\caption{Distribution of BAO uncertainties for the mock measurements (blue) versus two populations of Monte Carlo (MC) simulated correlation functions. The MC simulations are based on the best-fit model and covariance matrix measured from DESI DR1. The distribution in black uses the model that was used to fit the data, which includes BAO broadening due to non-linear evolution, while the distribution in red uses the same model but without the BAO broadening effect. The vertical dashed lines represent the uncertainties measured from DESI DR1. This shows that the BAO broadening effect can fully explain the difference between the BAO uncertainties measured in mocks versus data.}
\label{fig:sigma_distrib_mc_test}
\end{figure*}

As noted above, the most relevant effect for BAO that is not modelled in our mocks is the broadening due to non-linear evolution. To understand the impact of this effect, we performed the test presented in \Cref{fig:sigma_distrib_mc_test}. We used the covariance matrix measured by \kplya from DESI DR1 to create two populations of Monte Carlo simulations of our correlation functions following the method in \Cref{app:monte_carlo}. The first population is based on the best-fit model to DESI DR1, which includes the BAO broadening effect. This broadening is modelled by adding Gaussian smoothing to the peak component of the template power spectrum following \cite{Kirkby:2013}. The second population uses the same model but without the BAO broadening. We find that this second population matches very well the distribution of BAO uncertainties we measure from our mocks (\Cref{fig:sigma_distrib_mc_test}), while the first is shifted to larger uncertainties that match the actual measurement from DESI DR1. Therefore, at the level of DESI DR1, BAO broadening leads to a $\sim50\%$ increase in uncertainty for \apar, and $\sim25\%$ for \atrans. 

The model for this BAO broadening effect is based on Lagrangian perturbation theory \cite{Eisenstein:2007}, and is described in detail in \cite{Kirkby:2013}. It has been used in all \lyaf\ BAO analyses to date. Recently, \cite{Hadzhiyska2023} used \lyaf\ mocks based on N-body simulations to show that this model fits the BAO broadening effect very well. Therefore, while this feature is absent from our mocks, it is well understood and accounted for in \lyaf\ BAO analyses. Furthermore, \Cref{fig:sigma_distrib_mc_test} shows that the DESI DR1 uncertainties are consistent with the population of MC mocks produced from the model with BAO broadening (validated in \cite{Hadzhiyska2023}) and the covariance matrix (validated here). Nevertheless, our results here serve as a strong basis for moving towards more realistic mocks in future \lyaf\ analyses \cite[e.g.,][]{Sinigaglia:2021,Sinigaglia:2023,Hadzhiyska2023}.

Contamination due to metal absorption in the forest region is added both to our mocks and also modelled at the level of the correlation function. However, current-generation mocks have a relatively simplistic treatment for this metal absorption. We use \lya\ flux skewers and re-scale them such that resulting mocks reproduce the metal biases measured from real data (see \Cref{subsec:metal_tune}). This has two main weaknesses. The first was exposed in \Cref{subsec:res_zerr}, where we found that some of the metal biases are systematically shifted because of spurious correlations due to redshift errors. This means the metal tuning process cannot replicate realistic metal contamination unless we are able to properly model these spurious correlations. The second weakness comes from the fact that real metal contamination is associated with the circum-galactic and intra-galactic medium, rather than the IGM \cite{Perez-Rafols2023}. This means our use of the \lya\ flux skewers which trace the IGM only provides a very rough approximation for the metal contamination. A more realistic approach would be to draw metal lines from the peaks of the matter field, similar to how quasars and DLAs are drawn \cite{Farr:2020}. In terms of BAO analyses, metal contamination is most relevant due to the SiII$(1260)$ line which leads to the metal peak observed at $\sim103$\hMpc, close to where the BAO peak is located. However, we find that our model is able to tell the two apart both in the mocks here and in the DESI DR1 data (\kplya). Neither \apar\ nor \atrans\ is correlated with the bias parameter of this metal line, indicating that our measurement of the BAO position is not sensitive to differences between how this metal line is added in mocks versus how it appears in real data.

While our mocks contain the major \lyaf\ contaminants, there are a few less important effects that appear in real data but are not included here. These are:
\begin{itemize}
    \item \textbf{CIV contamination:} The auto-correlation of CIV absorption in the forest region has been modelled in previous \lyaf\ BAO analyses, but not detected at a significant level \cite{duMasdesBourboux:2020,DESI2024.IV.KP6}.

    \item \textbf{Transverse proximity effect:} We expect quasar radiation to have a significant impact on their surrounding environment, by increasing the ionization rate. This effect appears on small scales in the \lya-QSO cross-correlation and has been modelled in previous analyses using a simple model proposed by \cite{FontRibera2013}.

    \item \textbf{Correlated sky residuals:} Spectra from fibers in the same DESI petal \cite{Guy:2023} have correlated noise introduced by the data reduction pipeline. This has been studied in \cite{KP6s5-Guy}, and a simple model was proposed that can accurately account for this contamination. This model is included in the analysis of \kplya.

    \item \textbf{UV background fluctuations:} Fluctuations in the ionizing UV background can make the \lya\ bias and RSD parameters scale dependent \cite{Pontzen:2014,GontchoAGontcho:2014}. This has been modelled following \cite{GontchoAGontcho:2014} in previous \lyaf\ BAO analyses, but not detected at a significant level.
\end{itemize}
We note that none of these effects impact BAO measurements, and all have been tested as part of the blinded analysis performed in \kplya.

None of the limitations discussed here significantly affect our ability to use the mocks presented in this work to validate the DESI DR1 \lya\ BAO measurement. However, based on our results here, we have identified a few priorities for improving the next generation of synthetic data sets. These include going beyond the log-normal approximation in order to simulate non-linear broadening of the BAO peak and realistic quasar clustering down to small scales \cite[e.g.,][]{Sinigaglia:2021,Sinigaglia:2023,Hadzhiyska2023}, improving the realism of how metal contamination is added in mocks \cite[e.g.,][]{Farr:2020}, and generating a much larger set of mocks to improve our ability to study covariance matrix estimates and validate our uncertainties with improved precision. 

%%%%%%%%%%%%%%%%%%%%%%%%%%%%%%%%%%%%%%%%%%%%%%%%%%%%%%%%%%%%%%%%%%%%%%%%%%%%%%%%%%%%%%%%%%%%%%%%5

\section{Summary}
\label{sec:conclusions}

The first year of data from the Dark Energy Spectroscopic Instrument (DESI) contains the largest set of quasar spectra ever observed. Lyman-$\alpha$ (\lya) forests measured from these spectra were used by \kplya to measure Baryon Acoustic Oscillations (BAO) at an effective redshift $z=2.33$ with unprecedented precision. In this work, we use synthetic data sets (mocks) to perform the analysis validation for this DESI data release 1 (DR1) \lya\ BAO measurement.

We use a set of 150 mocks generated from Gaussian random fields with quasar positions and the \lya\ transmitted flux field inferred from its log-normal transformation. 100 of these were generated using \lyacolore\ \cite{Ramirez-Perez:2022,Farr:2020}, and 50 were generated using the \saclay\ framework described in \cite{Etourneau:2023}. These initial mocks were then used to simulate DESI DR1 quasar catalogues, and generate realistic DESI spectra following \cite{Herrera-Alcantar:2024}. The spectra include the major \lyaf\ contaminants, such as metal absorption, damped \lya\ systems (DLAs), broad absorption lines (BALs), and redshift errors. We describe the process of creating the mock data sets in \Cref{sec:mocks}.

We present our results in \Cref{sec:results}, where we perform two types of studies. First, we stack the correlation functions from all mocks to obtain very high statistics measurements of the \lya\ correlations and use these to check for potential systematic biases affecting the measurement of the BAO position. The results of this study are presented in \Cref{fig:stack}, which shows that using our mocks we are able to obtain unbiased measurements of the BAO parameters \apar\ and \atrans. Secondly, we analyze each mock individually and study the statistics of the resulting population of BAO fits in \Cref{subsec:res_pop}. We find that mock BAO constraints are randomly scattered around the truth (\Cref{fig:mock_pop}), with an RMS that roughly matches the measured uncertainties on \apar\ and \atrans\ (\Cref{fig:sigma_distrib}). Finally, we use the pull distribution shown in \Cref{fig:pull_distrib} to validate the BAO uncertainties at the $\sim9\%$ level. However, as our mocks do not include the effect of BAO broadening due to non-linear evolution, the uncertainties we measure here are significantly tighter than in real data (see \Cref{sec:discussion}).

In our baseline analysis, each mock has its own estimate of the covariance matrix. However, these are very noisy as they are based on the covariance of correlations computed in $1028$ patches on the sky (see \Cref{subsec:cov_mat} and \Cref{fig:correlation_matrices}). In \Cref{subsec:res_cov}, we test these noisy estimates by performing another analysis of the mock population using the normalized covariance (correlation matrix) estimated from the set of all mocks, which is much less noisy. We find that the choice of covariance matrix estimate does not have a significant impact on BAO constraints.

We test the impact of redshift errors in \Cref{subsec:res_zerr}, in the scenario where they are allowed to impact the continuum fitting process following \cite{Youles:2022}. Redshift errors give rise to smearing of emission lines in the forest region, and due to the clustering of quasars, these continuum errors lead to spurious correlations in both the \lya\ auto and its cross-correlation with quasars \cite{Youles:2022}. We find that these spurious correlations can produce a small but detectable systematic bias in the measured BAO peak position. This bias was found to only affect the Alcock-Paczynski direction and not the isotropic BAO measurement. However, even the Alcock-Paczynski BAO measurement was only biased by less than a tenth of the DESI DR1 uncertainty. While our results suggest this effect is not significant for DESI DR1, they do emphasize the need to model these spurious correlations, so future analyses can correctly marginalize them.

Finally, in \Cref{sec:discussion} we discuss the applicability and limitations of the results presented here. The main limitations include the fact that we use log-normal mocks, which do not include the effect of BAO broadening and overpredict the small-scale quasar clustering. However, the large scales used for BAO analyses are not significantly impacted by this \cite{Hadzhiyska2023}, and \Cref{fig:sigma_distrib_mc_test} combined with tests performed in \cite{DESI2024.IV.KP6,Hadzhiyska2023} show that we understand and can accurately model the effect of BAO broadening in real data. Nevertheless, we expect that the work and results presented here will motivate further development of more realistic mock data sets of the \lyaf.

The work presented here was used to validate the DESI DR1 \lyaf\ BAO analysis, and motivate decisions made during that analysis (\kplya). This led to the tightest BAO constraints from large-scale structure at redshifts $z>2$. \kplya measured the expansion rate at redshift $z=2.33$ with $2\%$ precision, and the transverse comoving distance to that redshift with $2.4\%$ precision.

%%%%%%%%%%%%%%%%%%%%%%%%%%%%%%%%%%%%%%%%%%%%%%%%%%%%%%%%%%%%%%%%%%%%%%%%%%%%%%%%%%%%%%%%%%%%%%%%5

\section{Data Availability}
The data used in this work will be made public as part of DESI Data Release 1 (details in \url{https://data.desi.lbl.gov/doc/releases/}). The data points corresponding to the most relevant figures
in this paper will be available in a Zenodo\footnote{\url{https://zenodo.org}, the exact link will be given when ready.} repository when it is accepted for publication.

\acknowledgments

AC acknowledges support provided by NASA through the NASA Hubble Fellowship grant HST-HF2-51526.001-A awarded by the Space Telescope Science Institute, which is operated by the Association of Universities for Research in Astronomy, Incorporated, under NASA contract NAS5-26555. 
HKHA and AXGM acknowledge support from Dirección de Apoyo a la Investigación y al Posgrado, Universidad de Guanajuato, research Grant No. 65/2024 and CONAHCYT México under Grants No. 286897 and A1-S-17899.
CG is partially supported by the Spanish Ministry of Science and Innovation (MICINN) under grants PGC-2018-094773-B-C31 and SEV-2016-0588. IFAE is partially funded by the CERCA program of the Generalitat de Catalunya. 
PM acknowledges support from the United States Department of Energy, Office of High Energy Physics under Award Number DE-SC0011726.
AFR acknowledges financial support from the Spanish Ministry of Science and Innovation under the Ramon y Cajal program (RYC-2018-025210) and the PGC2021-123012NB-C41 project, and from the European Union's Horizon Europe research and innovation programme (COSMO-LYA, grant agreement 101044612).

In addition to the packages already mentioned above, we also acknowledge the use of the \texttt{numpy} \cite{Harris:2020}, \texttt{scipy} \cite{scipy:2020}, \texttt{astropy} \cite{astropy:2013,astropy:2018,astropy:2022}, \texttt{mpi4py} \cite{mpi4py}, \texttt{healpy} \cite{healpy}, \texttt{matplotlib} \cite{matplotlib}, \texttt{GetDist} \cite{Lewis:2019}, \texttt{numba} \cite{numba:2015}, and \texttt{fitsio}\footnote{\url{https://github.com/esheldon/fitsio}} packages.

This material is based upon work supported by the U.S. Department of Energy (DOE), Office of Science, Office of High-Energy Physics, under Contract No. DE–AC02–05CH11231, and by the National Energy Research Scientific Computing Center, a DOE Office of Science User Facility under the same contract. Additional support for DESI was provided by the U.S. National Science Foundation (NSF), Division of Astronomical Sciences under Contract No. AST-0950945 to the NSF’s National Optical-Infrared Astronomy Research Laboratory; the Science and Technology Facilities Council of the United Kingdom; the Gordon and Betty Moore Foundation; the Heising-Simons Foundation; the French Alternative Energies and Atomic Energy Commission (CEA); the National Council of Humanities, Science and Technology of Mexico (CONAHCYT); the Ministry of Science and Innovation of Spain (MICINN), and by the DESI Member Institutions: \url{https://www.desi.lbl.gov/collaborating-institutions}. Any opinions, findings, and conclusions or recommendations expressed in this material are those of the author(s) and do not necessarily reflect the views of the U. S. National Science Foundation, the U. S. Department of Energy, or any of the listed funding agencies.

The authors are honored to be permitted to conduct scientific research on Iolkam Du’ag (Kitt Peak), a mountain with particular significance to the Tohono O’odham Nation.

\bibliographystyle{JHEP}
\bibliography{main,DESI2024}

\appendix

\section{Tests with Monte Carlo simulations}
\label{app:monte_carlo}

To gain further insight into our results from \Cref{subsec:res_pop}, we used populations of Monte Carlo (MC) simulations of our correlation functions following \cite{Cuceu:2020}. We start with the covariance matrix measured from the stack of 100 \lyacolore\ mocks and re-scaled to match the variance of one mock (multiplied by 100). We have also performed this test using the covariance matrix of one of the mocks and found similar results. Given a covariance matrix $C$ and correlation function $\xi$, we generate samples from the multivariate normal distribution with mean $\xi$ and covariance $C$. These samples are given by:
\begin{equation}
    \Tilde{\xi} = \xi + A \Vec{y},
\end{equation}
where the matrix $A$ is given by the Cholesky decomposition $C=AA^T$, $\Vec{y}$ is a vector of N independent standard normal variates, and N is the size of C. These MC simulations of the correlation function are then fit using the same model from \Cref{subsec:model}.

\begin{figure}
\centering
\includegraphics[width=0.7\linewidth]{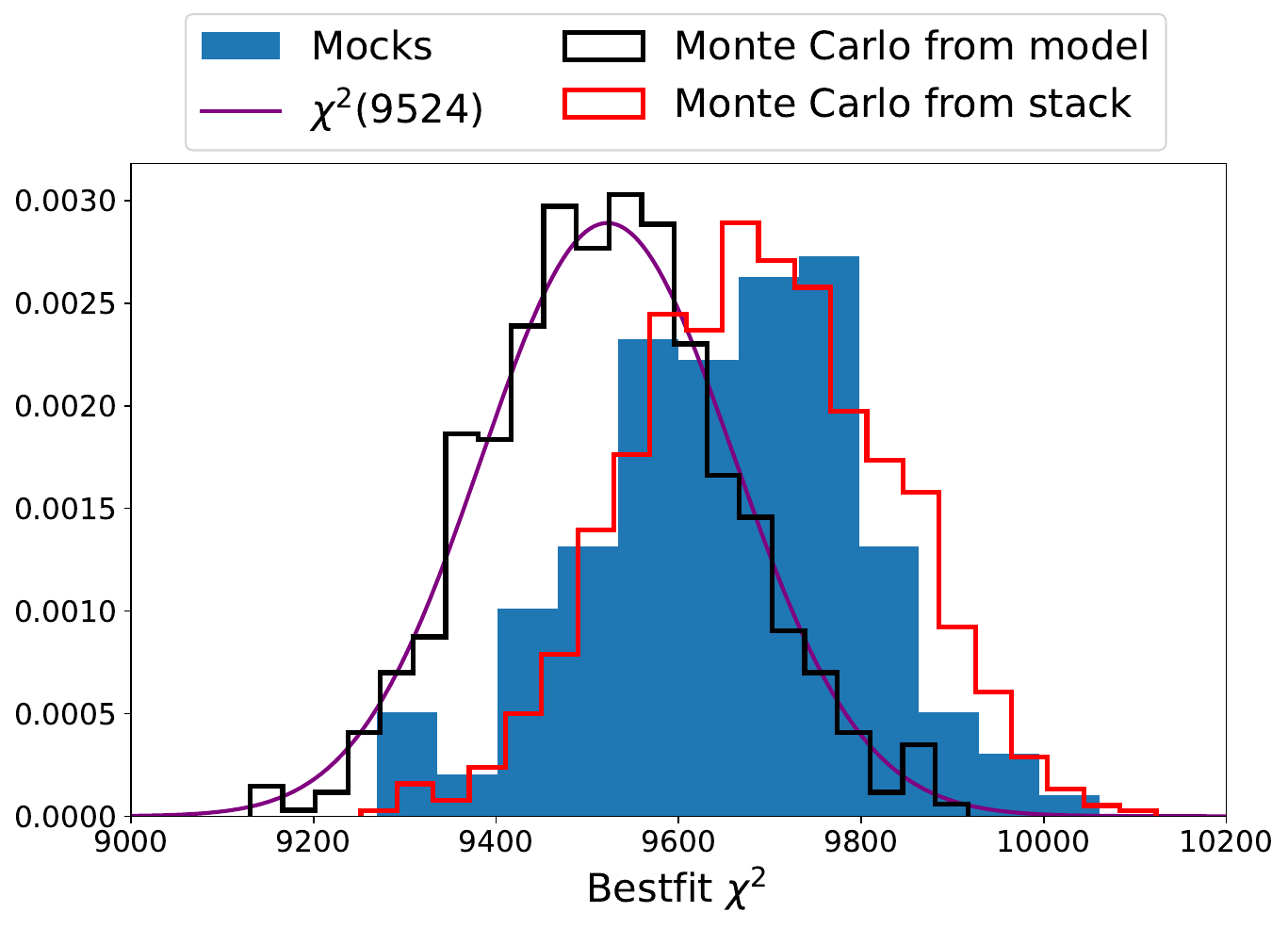}
\caption{Histograms of the best-fit $\chi^2$ statistic. We compare the distribution measured from mocks (blue), with two distributions obtained from Monte Carlo (MC) simulations of the correlation function. The first set of MC simulations was generated based on the stack of measured \lyacolore\ correlations in mocks (red), while the second was generated based on the best-fit model to the stack of 100 \lyacolore\ mocks (black). The distribution of MC simulations based on the model is consistent with the expected $\chi^2$ distribution with $9540-16=9524$ degrees of freedom, while the other two distributions are shifted to larger $\chi^2$ values. This indicates that the shift in mock best-fit $\chi^2$ values is caused by the model failing to accurately fit the mock correlation functions.}
\label{fig:mc_chisq_test}
\end{figure}

We generated two populations of MC simulations for our test. The first was created starting with the vector $\xi$ given by the stacked correlation functions from 100 \lyacolore\ mocks, $\xi_\mathrm{stack}$. The second was created starting from the best-fit model to the stack of 100 \lyacolore\ mocks, $\xi(\Vec{\theta}_\mathrm{best})$, where $\Vec{\theta}_\mathrm{best}$ represents the best-fit parameters. The two populations are then fit with the same model, and their resulting best-fit $\chi^2$ distributions are shown in \Cref{fig:mc_chisq_test}.

We find that the population based on the best-fit model matches the expected $\chi^2$ distribution with $9540-16=9524$ degrees of freedom very well. On the other hand, the population created from the stacked correlation functions is shifted to larger $\chi^2$ values and matches the observed $\chi^2$ distribution of our mocks. The covariance matrix used for these tests is correct by construction (i.e. the noise was generated with this covariance), and therefore, deviations from the expected $\chi^2$ distribution are caused by the inability of the model to fit the input $\xi$. This shows that the shift in the $\chi^2$ distribution measured from mocks is caused by our model failing to fit the measured correlation functions across the entire range of scales considered here ($10<r<180$\hMpc). We discuss this further in \Cref{subsec:res_pop}.

%%%%%%%%%%%%%%%%%%%%%%%%%%%%%%%%%%%%%%%%%%%%%%%%%%%%%%%%%%%%%%%%%%%%%%%%%%%%%%%%%%%%%%%%%%%%%%%%5

\section{Tests of the fiducial cosmology}
\label{app:fid_cosmo}

In the DESI DR1 BAO analysis, a fiducial cosmology based on the Planck 2018 results \cite{Planck:2020} is used to transform redshifts and angles to co-moving separations and as a template for the model. Here, we wish to validate that our constraints on $D_{\rm H}/r_{\rm d}$ and $D_{\rm M}/r_{\rm d}$ are independent of our choice of fiducial cosmology. To do this, we use a set of 20 \lyacolore\ DESI Y5 mocks, created using the method described in \Cref{sec:mocks}, but mirroring the footprint and exposure time (4000s) of the expected DESI Y5 survey. For simplicity, we use mocks that do not contain any of the contaminants discussed above (e.g., metal absorption, HCDs, etc.), but that do go through the same continuum fitting procedure described in section \ref{subsec:deltas}. We only use the \lyalyalyalya\ and \lyalyaqso\ correlations for this test.

\begin{figure}
\centering
\includegraphics[width=0.85\linewidth]{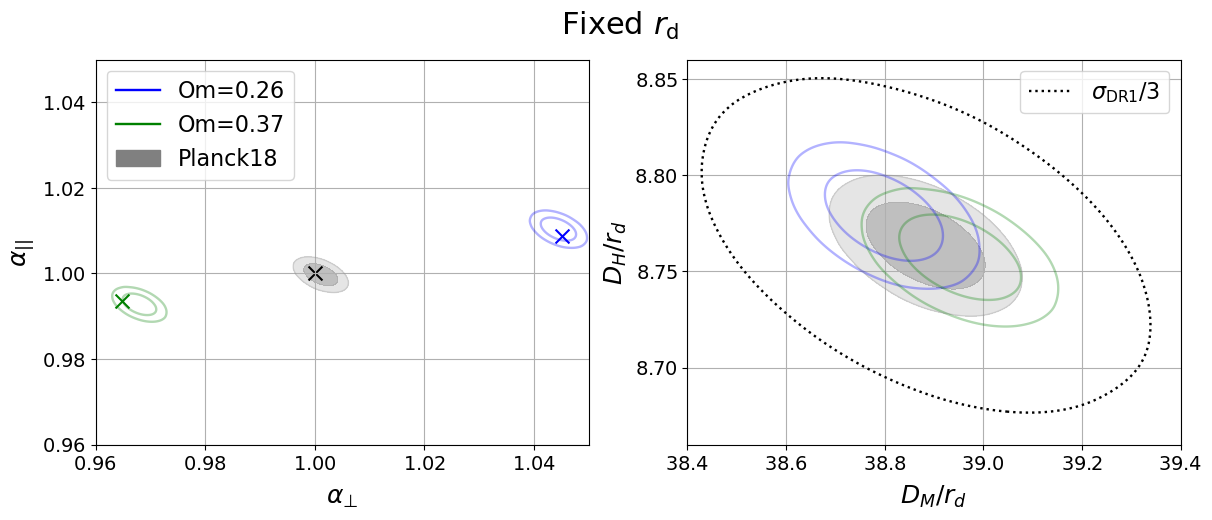}
\caption{(left) Scale parameters obtained from measurements with the three fiducial cosmologies with fixed $r_{\rm d}$ and different $\Omega_m$ and $h$ values. We also include crosses to mark the expected positions of the scale parameters, based on the ratio of their template BAO to that of the template used to create the mocks (Planck 2015). (right) Measured BAO distances obtained by multiplying the scale parameters with the template BAO position. This shows we are able to recover the true BAO position independent of the cosmology used to compute comoving coordinates.}
\label{fig:fix_rd}
\end{figure}

\begin{figure}
\centering
\includegraphics[width=0.85\linewidth]{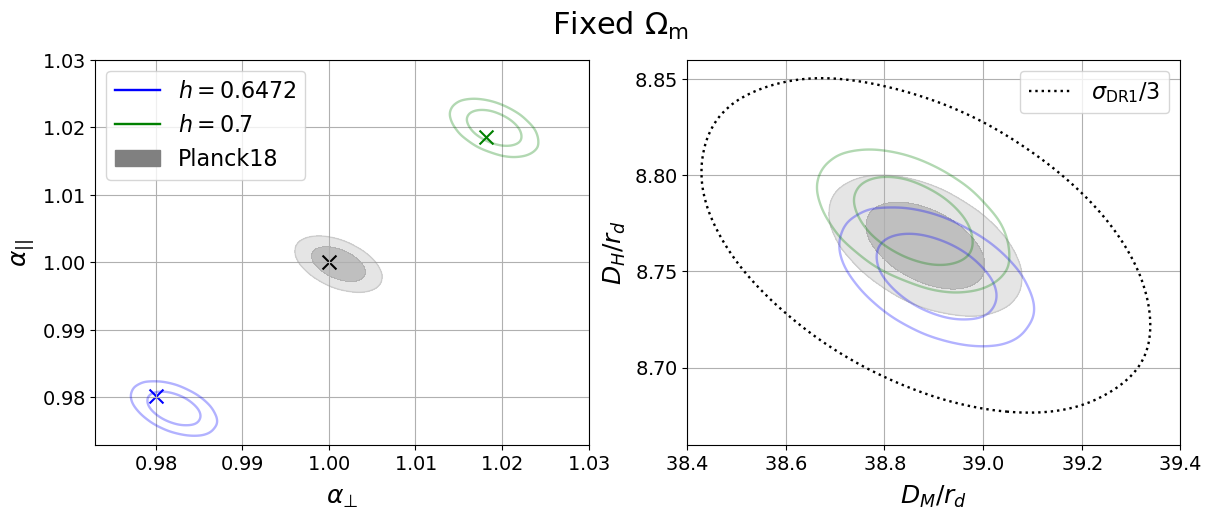}
\caption{Similar to \Cref{fig:fix_rd}, but using three fiducial cosmologies with fixed coordinate transformation (i.e. $\Omega_{\rm m}$), and different $r_d$ values. This shows that we are able to recover the true BAO position independent of the cosmology used to create the template.}
\label{fig:fix_om}
\end{figure}

To perform this test we use five cosmologies: Planck 2018 results \cite[Column 5 of Table 2 in][]{Planck:2020}, and four alternative cosmologies. Note that these are all different from the true cosmology used to create the \lyacolore\ mocks used here (see \Cref{subsec:mock_boxes}). We highlight this by showing the truth (as crosses) in our results in \Cref{fig:fix_rd,fig:fix_om}. In the first two of the four alternative cosmologies, we fix the sound horizon $r_{\rm d}$ and physical matter density $\Omega_{\rm m}h^2$ and vary $\Omega_{\rm m}$ and $h$ to the values given in \Cref{tab:alt_cosmo}. By fixing the sound horizon and physical matter density, which are both very well constrained from CMB anisotropies, we ensure the shape of our template power spectrum does not change. Thus we test the assumption that our BAO constraints are independent of the cosmology we use to make our coordinate transformation.

In the second set of alternative cosmologies, we instead fix $\Omega_{\rm m}$, while varying the values of $\Omega_bh^2$, $\Omega_ch^2$ and $h$. This changes $r_{\rm d}$ and $\Omega_{\rm m}h^2$, and therefore the shape of the template matter power spectrum, without affecting the coordinate transformation. The relevant values for this set are also given in \Cref{tab:alt_cosmo}. We choose values of $h=0.6472$ and $0.70$, which produces an $\pm$8\% change in $\Omega_{\rm m}h^2$ - roughly 10 times the error on the measurement from CMB anisotropies in Planck (2018). We note that in practice we do not expect such extreme differences between our template and the truth.
%\ref{eq:BAO}

\begin{table}[]
    \centering
    \begin{tabular}{cccccc}
        \hline
        \textbf{Parameter} & \textbf{Planck18} & \multicolumn{2}{c}{\textbf{Fix $r_{\rm d}$}} & \multicolumn{2}{c}{\textbf{Fix $\Omega_{\rm m}$}} \\
        \hline
        \hline
        $\Omega_{\rm m}$  & 0.315 & \multicolumn{1}{c|}{0.26} & \multicolumn{1}{c}{0.37} & \multicolumn{1}{c|}{0.315} & 0.315 \\
        $h$ & 0.6736 & \multicolumn{1}{c|}{0.7415} & \multicolumn{1}{c}{0.6216} & \multicolumn{1}{c|}{0.6472} & 0.7 \\
        $\Omega_{\rm m}h^2$ & 0.14297 & \multicolumn{1}{c|}{0.1429} & \multicolumn{1}{c}{0.1429} & \multicolumn{1}{c|}{0.1319} & 0.1545 \\
        %\hline
        %$\Omega_{\rm c}h^2$ & 0.12 & \multicolumn{1}{c|}{0.12} & \multicolumn{1}{c}{0.12} & \multicolumn{1}{c|}{0.109} & 0.132 \\
        %\hline
        \hline
        %\hline
        $r_{\rm d} \; [\rm Mpc]$ & 147.08 & \multicolumn{1}{c|}{147.08} & \multicolumn{1}{c}{147.08} & \multicolumn{1}{c|}{150.09} & 144.16 \\
        %\hline
        $D_{\rm H}(z_{\rm eff}=2.3) [\rm Mpc]$ & 1289.3 & \multicolumn{1}{c|}{1278.0} & \multicolumn{1}{c}{1297.5} & \multicolumn{1}{c|}{1342.0} & 1240.5 \\
        %\hline
        $D_{\rm M}(z_{\rm eff}=2.3) [\rm Mpc]$ & 5712.8 & \multicolumn{1}{c|}{5463.7} & \multicolumn{1}{c}{5918.8} & \multicolumn{1}{c|}{5946.0} & 5496.8 \\
        %\hline
        $D_{\rm H}(z_{\rm eff}=2.3)/r_{\rm d}$ & 8.77 & \multicolumn{1}{c|}{8.69} & \multicolumn{1}{c}{8.82} & \multicolumn{1}{c|}{8.94} & 8.61 \\
        %\hline
        $D_{\rm M}(z_{\rm eff}=2.3)/r_{\rm d}$ & 38.84 & \multicolumn{1}{c|}{37.14} & \multicolumn{1}{c}{40.24} & \multicolumn{1}{c|}{39.62} & 38.13 \\
        %\hline
        \hline
    \end{tabular}
    \caption{Main parameters of the cosmologies we used to test the BAO dependence on the choice of fiducial cosmology. Planck18 is the cosmology used in the BAO analysis on data, the two "Fix $r_d$" cosmologies change the coordinate transformation without changing the model template, and the "Fix $\Omega_{\rm m}$" cosmologies do the opposite. The first three rows contain primary parameters that we vary, and the last 5 are derived from these.}
    \label{tab:alt_cosmo}
\end{table}

\begin{table}[]
    %\centering
    \resizebox{\linewidth}{!}{
    \begin{tabular}{cccccc}
        \hline
        \textbf{Result} & \textbf{Planck18} & \multicolumn{2}{c}{\textbf{Fix $r_{\rm d}$}} & \multicolumn{2}{c}{\textbf{Fix $\Omega_{\rm m}$}} \\
        \hline
        \hline
         &  & \multicolumn{1}{c|}{$\Omega_{\rm m}=0.26$} & \multicolumn{1}{c}{$\Omega_{\rm m}=0.37$} & \multicolumn{1}{c|}{$h=0.6472$} & $h=0.7$ \\
        \hline
        $\alpha_\parallel$ & 0.9997$\pm$0.0017 & \multicolumn{1}{c|}{1.0103$\pm$0.0018} & \multicolumn{1}{c}{0.9928$\pm$0.0017} & \multicolumn{1}{c|}{0.9783$\pm$0.0016} & 1.0199$\pm$0.0018 \\
        $\alpha_\perp$ & 1.0011$\pm$0.0021 & \multicolumn{1}{c|}{1.0445$\pm$0.0021} & \multicolumn{1}{c}{0.9680$\pm$0.0020} & \multicolumn{1}{c|}{0.9821$\pm$0.0020} & 1.0191$\pm$0.0021 \\
        %\hline
        $D_{\rm H}/r_{\rm d}$ &  8.763$\pm$0.015 & \multicolumn{1}{c|}{8.779$\pm$0.016} & \multicolumn{1}{c}{8.757$\pm$0.015} & \multicolumn{1}{c|}{8.747$\pm$0.015} & 8.776$\pm$0.015 \\
        %\hline
        $D_{\rm M}/r_{\rm d}$ & 38.883$\pm$0.080 & \multicolumn{1}{c|}{38.799$\pm$0.079} & \multicolumn{1}{c}{38.953$\pm$0.082} & \multicolumn{1}{c|}{38.906$\pm$0.080} & 38.859$\pm$0.080 \\
        \hline
        $A_{\rm BAO}$  & 0.976$\pm$0.011 & \multicolumn{1}{c|}{1.078$\pm$0.012} & \multicolumn{1}{c}{0.885$\pm$0.011} & \multicolumn{1}{c|}{0.8037$\pm$0.0095} & 1.171$\pm$0.014 \\
        
    \end{tabular} }
    \caption{Measured scale parameters for each fiducial cosmology, along with the corresponding $D_{\rm H}/r_{\rm d}$ and $D_{\rm M}/r_{\rm d}$ values, which are consistent independent of the fiducial cosmology used. We also show the measured BAO amplitude for each case.}
    \label{tab:alt_cosmo_results}
\end{table}

The BAO scale parameters ($\alpha_\parallel$,$\alpha_\perp$) are defined in \Cref{eq:BAO}, as the ratio of the measured BAO peak position to that of the template used in the analysis. Therefore, we expect them to shift when fitting our mock datasets with different cosmologies. The results are summarized in \Cref{tab:alt_cosmo_results}, and the recovered scale parameters from each of our fits are shown on the left-hand panel of \Cref{fig:fix_rd,fig:fix_om}. The crosses in each case correspond to the predicted scale parameter locations, calculated from the ratio $[D_{\rm H}/r_{\rm d}]_{\rm true}/[D_{\rm H}/r_{\rm d}]_{\rm X}$, where X stands for the alternative cosmology in question. These plots show the large shifts introduced by changing the fiducial cosmology, and how well we can recover the expected values. The expected values (crosses) are within the $68\%$ contour in all cases.

In the right panel of \Cref{fig:fix_rd,fig:fix_om}, we show the recovered BAO distances, obtained by multiplying the measured scale parameter (contour in the left-hand plot) by the template BAO position (table \ref{tab:alt_cosmo}). Also highlighted on this plot is the DESI DR1 $\sigma/3$, the limit given to variations around the baseline analysis in \kplya. These figures show that we are able to recover the correct $D_{\rm H}/r_{\rm d}$ and $D_{\rm M}/r_{\rm d}$ values to within $1/10$ of the DESI DR1 uncertainty. This also corresponds to the level at which we can trust the current generation of mocks given the results in \Cref{fig:stack}. Therefore, we conclude that at the level of DESI DR1, we do not detect a significant systematic offset due to the input fiducial cosmology.

In \Cref{tab:alt_cosmo_results}, we also show the measured BAO amplitude $A_{\rm BAO}$, which we treat as a free parameter for this test. This parameter modifies \Cref{eq:xi_sum} such that:
\begin{equation}
    \xi(r_{||},r_\bot) = \hat\xi_\mathrm{s}(r_{||}, r_\bot) + A_{\rm BAO} \hat\xi_\mathrm{p}(\alpha_{||} r_{||}, \alpha_\bot r_\bot),
\end{equation}
and therefore it fits for differences between the template and the measured amplitude of the BAO peak. Note that this is fixed to $A_{\rm BAO}=1$ for the rest of this work, and also in \kplya. As shown in \Cref{tab:alt_cosmo_results}, we find that the recovered value for this parameter changes depending on the fiducial cosmology. For the cases where we vary $\Omega_m h^2$, the ratio $\Omega_c h^2$/$\Omega_b h^2$ also varies, and therefore the BAO amplitude as well. For the cases where we change $\Omega_m$, it is not as straightforward. It could be explained by the fact that f$\sigma_8$ changes, which affects the value of $b_{Ly\alpha}$ and leads to different BAO amplitudes, as $A_{\rm BAO}$ is correlated with $b_{Ly\alpha}$. However, as we do not use this parameter to constrain cosmology, we defer an in-depth study of its reliance on the fiducial cosmology to future work.

\section{Author Affiliations}
\label{sec:affiliations}

\noindent \hangindent=.5cm $^{1}${Center for Cosmology and AstroParticle Physics, The Ohio State University, 191 West Woodruff Avenue, Columbus, OH 43210, USA}

\noindent \hangindent=.5cm $^{2}${Departamento de F\'{i}sica, Universidad de Guanajuato - DCI, C.P. 37150, Leon, Guanajuato, M\'{e}xico}

\noindent \hangindent=.5cm $^{3}${Institut de F\'{i}sica d’Altes Energies (IFAE), The Barcelona Institute of Science and Technology, Campus UAB, 08193 Bellaterra Barcelona, Spain}

\noindent \hangindent=.5cm $^{4}${Department of Astronomy, The Ohio State University, 4055 McPherson Laboratory, 140 W 18th Avenue, Columbus, OH 43210, USA}

\noindent \hangindent=.5cm $^{5}${Lawrence Berkeley National Laboratory, 1 Cyclotron Road, Berkeley, CA 94720, USA}

\noindent \hangindent=.5cm $^{6}${Consejo Nacional de Ciencia y Tecnolog\'{\i}a, Av. Insurgentes Sur 1582. Colonia Cr\'{e}dito Constructor, Del. Benito Ju\'{a}rez C.P. 03940, M\'{e}xico D.F. M\'{e}xico}

\noindent \hangindent=.5cm $^{7}${IRFU, CEA, Universit\'{e} Paris-Saclay, F-91191 Gif-sur-Yvette, France}

\noindent \hangindent=.5cm $^{8}${Physics Dept., Boston University, 590 Commonwealth Avenue, Boston, MA 02215, USA}

\noindent \hangindent=.5cm $^{9}${Department of Physics and Astronomy, University of California, Irvine, 92697, USA}

\noindent \hangindent=.5cm $^{10}${Department of Physics \& Astronomy, University College London, Gower Street, London, WC1E 6BT, UK}

\noindent \hangindent=.5cm $^{11}${Instituto de F\'{\i}sica, Universidad Nacional Aut\'{o}noma de M\'{e}xico,  Cd. de M\'{e}xico  C.P. 04510,  M\'{e}xico}

\noindent \hangindent=.5cm $^{12}${Kavli Institute for Particle Astrophysics and Cosmology, Stanford University, Menlo Park, CA 94305, USA}

\noindent \hangindent=.5cm $^{13}${SLAC National Accelerator Laboratory, Menlo Park, CA 94305, USA}

\noindent \hangindent=.5cm $^{14}${University of California, Berkeley, 110 Sproul Hall \#5800 Berkeley, CA 94720, USA}

\noindent \hangindent=.5cm $^{15}${Departamento de F\'isica, Universidad de los Andes, Cra. 1 No. 18A-10, Edificio Ip, CP 111711, Bogot\'a, Colombia}

\noindent \hangindent=.5cm $^{16}${Observatorio Astron\'omico, Universidad de los Andes, Cra. 1 No. 18A-10, Edificio H, CP 111711 Bogot\'a, Colombia}

\noindent \hangindent=.5cm $^{17}${Institut d'Estudis Espacials de Catalunya (IEEC), 08034 Barcelona, Spain}

\noindent \hangindent=.5cm $^{18}${Institute of Cosmology and Gravitation, University of Portsmouth, Dennis Sciama Building, Portsmouth, PO1 3FX, UK}

\noindent \hangindent=.5cm $^{19}${Institute of Space Sciences, ICE-CSIC, Campus UAB, Carrer de Can Magrans s/n, 08913 Bellaterra, Barcelona, Spain}

\noindent \hangindent=.5cm $^{20}${Fermi National Accelerator Laboratory, PO Box 500, Batavia, IL 60510, USA}

\noindent \hangindent=.5cm $^{21}${Department of Physics, The Ohio State University, 191 West Woodruff Avenue, Columbus, OH 43210, USA}

\noindent \hangindent=.5cm $^{22}${School of Mathematics and Physics, University of Queensland, 4072, Australia}

\noindent \hangindent=.5cm $^{23}${Sorbonne Universit\'{e}, CNRS/IN2P3, Laboratoire de Physique Nucl\'{e}aire et de Hautes Energies (LPNHE), FR-75005 Paris, France}

\noindent \hangindent=.5cm $^{24}${Departament de F\'{i}sica, Serra H\'{u}nter, Universitat Aut\`{o}noma de Barcelona, 08193 Bellaterra (Barcelona), Spain}

\noindent \hangindent=.5cm $^{25}${NSF NOIRLab, 950 N. Cherry Ave., Tucson, AZ 85719, USA}

\noindent \hangindent=.5cm $^{26}${Instituci\'{o} Catalana de Recerca i Estudis Avan\c{c}ats, Passeig de Llu\'{\i}s Companys, 23, 08010 Barcelona, Spain}

\noindent \hangindent=.5cm $^{27}${Department of Physics and Astronomy, Siena College, 515 Loudon Road, Loudonville, NY 12211, USA}

\noindent \hangindent=.5cm $^{28}${Department of Physics \& Astronomy, University  of Wyoming, 1000 E. University, Dept.~3905, Laramie, WY 82071, USA}

\noindent \hangindent=.5cm $^{29}${Instituto Avanzado de Cosmolog\'{\i}a A.~C., San Marcos 11 - Atenas 202. Magdalena Contreras, 10720. Ciudad de M\'{e}xico, M\'{e}xico}

\noindent \hangindent=.5cm $^{30}${Department of Physics and Astronomy, University of Waterloo, 200 University Ave W, Waterloo, ON N2L 3G1, Canada}

\noindent \hangindent=.5cm $^{31}${Perimeter Institute for Theoretical Physics, 31 Caroline St. North, Waterloo, ON N2L 2Y5, Canada}

\noindent \hangindent=.5cm $^{32}${Waterloo Centre for Astrophysics, University of Waterloo, 200 University Ave W, Waterloo, ON N2L 3G1, Canada}

\noindent \hangindent=.5cm $^{33}${Space Sciences Laboratory, University of California, Berkeley, 7 Gauss Way, Berkeley, CA  94720, USA}

\noindent \hangindent=.5cm $^{34}${Instituto de Astrof\'{i}sica de Andaluc\'{i}a (CSIC), Glorieta de la Astronom\'{i}a, s/n, E-18008 Granada, Spain}

\noindent \hangindent=.5cm $^{35}${Departament de F\'isica, EEBE, Universitat Polit\`ecnica de Catalunya, c/Eduard Maristany 10, 08930 Barcelona, Spain}

\noindent \hangindent=.5cm $^{36}${Aix Marseille Univ, CNRS/IN2P3, CPPM, Marseille, France}

\noindent \hangindent=.5cm $^{37}${Universit\'{e} Clermont-Auvergne, CNRS, LPCA, 63000 Clermont-Ferrand, France}

\noindent \hangindent=.5cm $^{38}${Department of Physics, Kansas State University, 116 Cardwell Hall, Manhattan, KS 66506, USA}

\noindent \hangindent=.5cm $^{39}${Department of Physics and Astronomy, Sejong University, Seoul, 143-747, Korea}

\noindent \hangindent=.5cm $^{40}${CIEMAT, Avenida Complutense 40, E-28040 Madrid, Spain}

\noindent \hangindent=.5cm $^{41}${Department of Physics, University of Michigan, Ann Arbor, MI 48109, USA}

\noindent \hangindent=.5cm $^{42}${University of Michigan, Ann Arbor, MI 48109, USA}

\noindent \hangindent=.5cm $^{43}${Department of Physics \& Astronomy, Ohio University, Athens, OH 45701, USA}

\noindent \hangindent=.5cm $^{44}${Excellence Cluster ORIGINS, Boltzmannstrasse 2, D-85748 Garching, Germany}

\noindent \hangindent=.5cm $^{45}${University Observatory, Faculty of Physics, Ludwig-Maximilians-Universit\"{a}t, Scheinerstr. 1, 81677 M\"{u}nchen, Germany}

\noindent \hangindent=.5cm $^{46}${National Astronomical Observatories, Chinese Academy of Sciences, A20 Datun Rd., Chaoyang District, Beijing, 100012, P.R. China}

% Bibliography

%% [A] Recommended: using JHEP.bst file

\end{document}